\documentclass[reprint,footinbib,superscriptaddress,floatfix, amsmath,amssymb, aps,prl, longbibliography]{revtex4-2}

\usepackage{graphicx}  
\usepackage{dcolumn}   
\usepackage{bm}        
\usepackage{amssymb}   
\usepackage{amsmath}
\usepackage{xcolor}
\usepackage{color}
\usepackage[colorlinks=true, pdfstartview=FitV, linkcolor=blue, citecolor=blue, urlcolor=blue]{hyperref} 
\usepackage{epstopdf}
\usepackage{siunitx} 
\usepackage{tabularx} 
\usepackage{xspace}
\usepackage{braket}
\usepackage{color}
\usepackage{setspace}
\usepackage{mdframed}
\usepackage{framed}
\usepackage{multirow}
\usepackage{setspace}
\usepackage[export]{adjustbox}
\usepackage{varwidth}
\usepackage{float}

\usepackage{lipsum}
\usepackage[version=3]{mhchem}
\usepackage{color, colortbl}
\usepackage[caption=false]{subfig}
\usepackage{booktabs}
\usepackage{array}
\usepackage{tabularx}
\usepackage{soul}
\definecolor{beaublue}{rgb}{0.74, 0.83, 0.9}

\newcommand{\affilETH}{Laboratory for Solid State Physics, ETH Z\"{u}rich, 8093 Z\"urich, Switzerland}
\newcommand{\affilQC}{Quantum Center, ETH Z\"{u}rich, 8093 Z\"{u}rich, Switzerland}
\newcommand{\affilKonst}{Department of Physics, University of Konstanz, 78464 Konstanz, Germany}

\def \p {\hat{p}}
\def \q {\hat{q}}
\def \hIx {\hat{I}_x}
\def \hIy {\hat{I}_y}
\def \hIz {\hat{I}_z}
\def \wo {\omega_\mathrm{0}}
\def \wdd {\omega_\mathrm{d}}
\def \wl {\omega_\mathrm{L}}
\def \cwd {\cos(\wdd t)}
\def \swd {\sin(\wdd t)}

\def \Gm {\Gamma_\mathrm{m}}

\def \lb {\left<}
\def \rb {\right>}

\def \gp {1/T_2}
\def \kappa {1/T_1}

\def \dhIz {\dot{\hat{I}}_z}
\def \Ix {I_x}
\def \Iy {I_y}
\def \Iz {I_z}

\def \dIz {\dot{I}_z}
\def \dw {\delta\omega}

\def \dFo {\delta F_0}
\def \dGm {\delta\Gm}

\begin{document}

\title{Near-resonant nuclear spin detection with high-frequency mechanical resonators}

\author{Diego A. Visani}
\affiliation{\affilETH}
\affiliation{\affilQC}
\author{Letizia Catalini}
\affiliation{\affilETH}
\affiliation{\affilQC}
\author{Christian L. Degen}
\affiliation{\affilETH}
\affiliation{\affilQC}
\author{Alexander Eichler}
\affiliation{\affilETH}
\affiliation{\affilQC}
\email[Corresponding author: ]{eichlera@ethz.ch}
\author{Javier del Pino}
\affiliation{Institute for Theoretical Physics, ETH Z{\"u}rich, 8093 Z{\"u}rich, Switzerland}
\affiliation{\affilKonst}

\begin{abstract}
Mechanical resonators operating in the high-frequency regime have become a versatile platform for fundamental and applied quantum research. Their exceptional properties, such as low mass and high quality factor, make them also very appealing for force sensing experiments. In this Letter, we propose a method for detecting and ultimately controlling nuclear spins by directly coupling them to high-frequency resonators via a magnetic field gradient. Dynamical backaction between the sensor and an ensemble of nuclear spins produces a shift in the sensor's resonance frequency, which can be measured to probe the spin ensemble. Based on analytical as well as numerical results, we predict that the method will allow nanoscale magnetic resonance imaging with a range of realistic devices. At the same time, this interaction paves the way for new manipulation techniques, similar to those employed in cavity optomechanics, enriching both the sensor's and the spin ensemble's features.
\end{abstract}

\maketitle

Magnetic resonance force microscopy (MRFM) is a method to achieve nanoscale magnetic resonance imaging (MRI)~\cite{Sidles_1991,Poggio_2010}. It relies on a mechanical sensor interacting via a magnetic field gradient with an ensemble of nuclear spins. The interaction creates signatures in the resonator oscillation that can be used to detect nuclear spins with high spatial resolution. Previous milestones include the imaging of virus particles with $5-\SI{10}{\nano\meter}$ resolution~\cite{Degen_2009}, Fourier-transform nanoscale MRI~\cite{Nichol_2013}, nuclear spin detection with a one-dimensional resolution below \SI{1}{\nano\meter}~\cite{Grob_2019}, and magnetic resonance diffraction with subangstrom precision~\cite{Haas_2022}. In all of these experiments, the excellent force sensitivity of the mechanical resonator is a key factor. The MRFM community is thus constantly searching for improved force sensors to reach new regimes of spin-mechanics interaction. 

Over the last decade, new classes of mechanical resonators made from strained materials showed promise as force sensors~\cite{Eichler_2022}. Today, these resonators come in a large variety of designs, including trampolines~\cite{Reinhardt_2016,Norte_2016}, membranes~\cite{Tsaturyan_2017,Rossi_2018,Reetz_2019}, strings~\cite{Ghadimi_2018,Beccari_2022,Gisler_2022}, polygons~\cite{Bereyhi_2022}, hierarchical structures~\cite{Fedorov_2020,Bereyhi_2022_NC}, and spider webs~\cite{Shin_2021}. Some of these resonators are massive enough to be seen by the naked eye, but their low dissipation nevertheless makes them excellent sensors, potentially on par with carbon nanotubes~\cite{Moser_2013} and nanowires~\cite{Rossi_2016,Nichol_2013}.
Using these strained resonators for nuclear spin detection requires novel scanning force geometries~\cite{Halg_2021} and transduction protocols~\cite{Kosata_2020}. At the same time, new experimental opportunities become feasible,  thanks to the ability of high-$Q$ mechanical systems to strongly interact with a wide array of quantum systems, such as nuclear spins, artificial atoms, and photonic resonators~\cite{Aspelmeyer_2014,RevModPhys.94.045005}.

\begin{figure}[h!]
    \centering
    \includegraphics[width=0.48\textwidth]{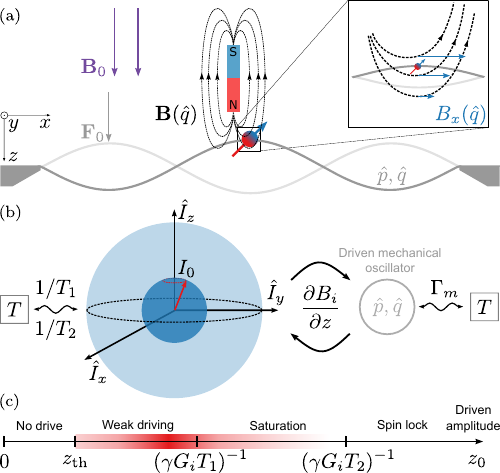}
    \caption{(a)~Proposed experiment: A spin ensemble is placed on a mechanical resonator moving within an inhomogeneous magnetic field generated by a nanoscale magnet. By driving the resonator, the spins experience an oscillating magnetic field $\textbf{B}(\q)$ with a component $B_x$ (inset). The spins act back on the resonator, producing a force that can be detected as a shift in the resonator frequency. (b)~We model the system as a spin ensemble (equilibrium polarization $I_\parallel=I_0$) interacting with a harmonic oscillator. Both spin ensemble and resonator are coupled to independent baths at temperature $T$, causing spin dephasing and decay with rates $\gp$ and $\kappa$, respectively, and resonator damping at a rate $\Gamma_m$. (c)~Illustration of the typical spin regimes according to driven mechanical amplitudes $z_0$. Here $z_\mathrm{th}$ denotes the thermal motional amplitude. The regime addressed in this work is highlighted in red.
    }
    \label{fig:fig1}
\end{figure}

In this work, we propose a protocol for nuclear spin detection based on the near-resonant interaction between a mechanical resonator and an ensemble of nuclear spins. In contrast to earlier ideas~\cite{Sidles_1992,Berman_2022}, our method is most efficient when the resonator is detuned from the spin Larmor frequency. The coupling then causes a shift of the mechanical resonance frequency which can be measured to quantify the spin states. Importantly, in contrast to the driven amplitude predicted in Ref.~\cite{Sidles_1992}, the response time to the frequency shift is not limited by the ringdown time. Our proposal suits the typical frequency range of strained resonators ($1-\SI{50}{\mega\hertz}$), but can also be applied to other state-of-the-art systems, such as graphene or carbon nanotube resonators ($\gtrsim \SI{100}{\mega\hertz}$). Our approach offers a radically simplified experimental apparatus, as it circumvents the need for spin inversion pulses and the related hardware. We also show that for realistic parameters, our method can attain single nuclear spin sensitivity, which would be a major milestone on the way towards spin-based quantum devices. Finally, it opens up a new field of spin manipulation via mechanical driving, drawing parallels to techniques used in cavity optomechanics~\cite{Aspelmeyer_2014, Lee_2016, Teissier_2017, MacQuarrie_2013}.

We consider a nuclear spin ensemble placed on a mechanical resonator, see Fig.~\ref{fig:fig1}(a). The ensemble comprises $N$ spins that interact with a normal mode of the resonator. The composite system can be described with the Zeeman-like Hamiltonian
\begin{equation}
    \mathcal{H}=-\hbar\gamma \hat{\textbf{I}}\cdot \textbf{B} + \mathcal{H}_m,
\end{equation}
where $\hbar$ is the reduced Planck constant, $\gamma$ the nuclear spin's gyromagnetic ratio, and $\textbf{B}$ the magnetic field at the spins' location. The spin ensemble operator has the three components $\hat{I}_i = \sum_{k=1}^N\hat{\sigma}_{i,k}/2$ with the spin-$\frac{1}{2}$ Pauli matrices $\hat{\sigma}_{i,k}$ for spin $k=\{1,\cdots,N\}$, and $i\in [x, y, z]$. We describe a single vibrational mode as a driven harmonic oscillator displacing along the $z$ axis governed by the Hamiltonian
\begin{equation}
    \mathcal{H}_m= \frac{\p^2}{2m}+\frac{1}{2}m\wo^2\q^2 - F_0\q\cos(\wdd t),
\end{equation}
where $\hat{q}$ is the $z$-position operator of the resonator at the position of the spin ensemble, $\hat{p}$ is the corresponding momentum operator, $m$ is the effective mass, $\wo$ is the resonance frequency, and $\wdd$ and $F_0$ are the frequency and strength of an applied force, respectively. If $\textbf{B}$ is inhomogeneous, the spins experience a position-dependent field $\textbf{B}(\q)$ as the mechanical resonator vibrates. To lowest order, we approximate this field as $\textbf{B}(\q)\approx \textbf{B}_0 + \textbf{G}\q$ with a constant component $\textbf{B}_0 = \textbf{B}(\q=0)$ and relevant field gradients $G_i=\partial B_i/\partial z$. The coherent spin-resonator dynamics therefore obey the Hamiltonian 
\begin{equation}
    \mathcal{H} \approx -\hbar\wl\hIz -\hbar\gamma\q \textbf{G}\cdot \hat{\textbf{I}}+\mathcal{H}_m,
\end{equation}
with the Larmor precession frequency $\wl=\gamma|B_0|$.

Any real system, in equilibrium with a thermal bath, experiences mechanical damping (rate $\Gm$), spin decay (longitudinal relaxation time $T_1$), and spin decoherence (transverse relaxation time $T_2$). We thus succinctly represent our system dynamics using the Heisenberg picture's dissipative equations of motion (EOM).
Driving the resonator to an oscillation amplitude $z_0$ well above its zero-point fluctuation amplitude $z_\mathrm{zpf}=\sqrt{\hbar/(2m\omega_0)}$,
we treat the mechanical resonator classically in the mean-field approximation. This allows us to approximate the averaged operators as the products of their respective averages, e.g. replacing $\braket{\hat{q}\hat{I}_i}$ by $\braket{\hat{q}}\braket{\hat{I}_i}$. These averages then evolve via time-dependent Bloch equations~\cite{Supplement}
\begin{align}
    \ddot{q} &=  -\wo^2q -\Gm\dot{q} + \frac{F_0}{m}\cwd + \frac{\hbar\gamma}{m}\textbf{G}\cdot \textbf{I},\label{eq:eom_q_fin}\\
	\dot{I}_{x,y}&=-\frac{1}{T_2}I_{x,y}\pm(\wl+\gamma qG_{z})I_{y,x}\mp\gamma qG_{y,x}I_{z}\label{eq:eom_Ixy_fin} ,\\
	\dIz &= \frac{1}{T_1}\left(I_0 - \Iz\right) - \gamma q\left(G_x\Iy - G_y\Ix\right),\label{eq:eom_Iz_fin}
\end{align}
where we dropped the $\lb...\rb$ and the hat notation. Here, $I_0$ denotes the Boltzmann polarization. In the limit $k_BT \gg \hbar\wl$, it simplifies to $I_0\approx N\hbar\wl/(4k_BT)$~\cite{niinikoski_2020}.
The thermomechanical fluctuations do not impact the average resonator's position $q$, leaving Eq.~\eqref{eq:eom_q_fin} explicitly independent of temperature $T$. Note that in our simplified model, the spins' decay (decoherence) time $T_1$ ($T_2$) is independent of temperature and magnetic field.

To simplify the treatment, we restrict ourselves to the regime where (i)~the spins' force on the resonator is weak compared to the driving force, $|\delta F| \ll F_0$, and (ii)~the spin-resonator coupling, parameterized by a Rabi frequency $\Omega_R = \gamma G_i z_0$, is weaker than dissipation, meaning $\Omega_R \ll \gp, \kappa$, see Fig.~\ref{fig:fig1}(c). In order to fulfill (ii), we select $z_0$ to be small and on the order of the thermal motion $z_\mathrm{th}$~\cite{Supplement}.
The conditions (i) and (ii) imply that we remain in the weak coupling limit, where the oscillation inside the field gradient $\textbf{G}$ excites a precessing spin polarization orthogonal to $\textbf{B}_0$ (i.e., $I_{x,y}\neq 0$), but does not lock the spins to the resonator frequency $\omega_0$. The backaction of the spins can be treated as a perturbation of the driven resonator oscillation at frequency $\wdd$. Note that, in the absence of spin locking, spin-spin interactions will lead to short $T_2$ times on the order of \SI{100}{\micro\second} for dense spin ensembles~\cite{Bloch_1946}.

The spin components $I_{x,y}$ exert a linear force back onto the resonator that we calculate via the Harmonic Balance method~\cite{Krack_2019, Kosata_2022_SP} as detailed in~\cite{Supplement}. The force involves a static component  $\dFo = \hbar\gamma I_0G_z/m$ that shifts the mechanical equilibrium, and two oscillating components, in-phase and quadrature. This \textit{dynamical backaction} loop causes a frequency shift $\dw$ (corresponding to a phase shift in the driven response) and a linewidth change $\dGm$~\cite{Supplement}: 
\begin{align}
\dw &= -g^2\left(\frac{\omega_+}{\omega_+^2 + \gp^2} + \frac{\omega_-}{\omega_-^2 + \gp^2 }\right),\label{eq:dyn_back_simple} \\
\dGm &= -g^2\left(\frac{\gp}{ \omega_+^2+\gp^2}-\frac{\gp}{\omega_-^2 + \gp^2}\right), \label{eq:dyn_back_simple_2}
\end{align}
where $\omega_{\pm}=\wl\pm\wdd$ and $g^2=\hbar\gamma^2 I_0\left(G_x^2+G_y^2\right)/(4m\wdd)$.

\begin{figure}[t]
    \centering
    \includegraphics[width=0.483\textwidth]{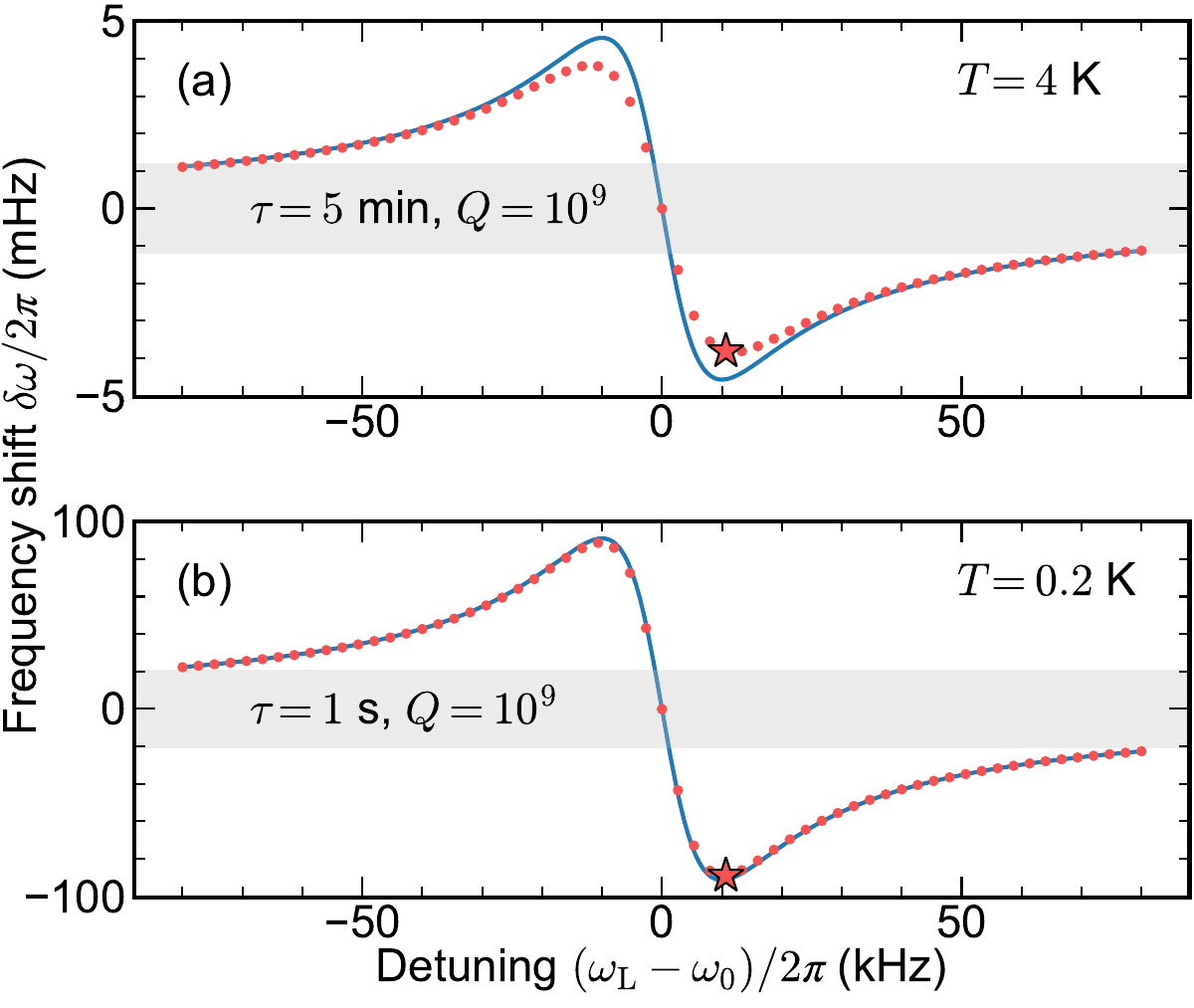}
    \caption{Frequency shift $\dw$ of a string resonator~\cite{Ghadimi_2018} calculated as a function of the detuning between the Larmor frequency $\wl$ and the mechanical frequency $\wo$. Analytical and numerical results are shown for temperatures of (a)~\SI{4}{\kelvin} and (b)~\SI{0.2}{\kelvin}. Blue lines correspond to Eq.~\eqref{eq:dyn_back_simple}, while red dots are calculated with an explicit Runge-Kutta method of order 8~\cite{Hairer_1993}. A maximal frequency shift in each case is shown by a red star.
    The shaded area represents the thermal frequency noise standard deviation for an integration time $\tau$~\cite{Allan_1972,Walls_1986}. 
    Common simulated parameters are $\wdd=\wo=2\pi\times\SI{5.5}{\mega\hertz}$, $G_x=G_y=\SI{2}{\mega\tesla/\meter}$, $G_z=\SI{1}{\mega\tesla/\meter}$, $m=\SI{2}{\pico\gram}$ and $N=10^6$. See Table~\ref{tab:fig_merit} for all device details.}
    \label{fig:Fig2}
\end{figure}

The results in Eq.~\eqref{eq:dyn_back_simple} and in Fig.~\ref{fig:Fig2} demonstrate that the equilibrium spin population $I_0$ alters the resonator properties through dynamical back-action: the force produced by the oscillating in-plane components $I_x$ and $I_y$ is \textit{delayed} with respect to the drive. The force is most pronounced when spins interact with the resonator faster than the resonator's reaction time ($\Gamma_m\ll\gp,\kappa$). This spin-mediated delayed force creates a closed feedback loop that modifies the resonator's stiffness (frequency) and  enhances or suppresses damping, depending on when the force peaks in relation to position or velocity. Importantly, the strongest spin-mechanics interaction occurs at a detuning $\wl\neq\wo$ that depends on $\gp$ and $\wl$, see Eq.~\eqref{eq:dyn_back_simple}. 
This is different from the early MRFM proposal that considered resonant coupling forces where $\wl=\wo$~\cite{Sidles_1992}, and from spin noise measurements in MRI~\cite{McCoy_1989,Gueron_1989}. Instead, the effect we rely on resembles dynamical back-action 
in e.g. cavity optomechanics~\cite{Braginsky_1967,Aspelmeyer_2014}.

For our spin sensing protocol, measuring the frequency shift $\dw$ is advantageous over measuring the linewidth broadening $\dGm$, as the frequency response rate is not intrinsically limited by the ringdown time~\cite{Albrecht_1991}. In Fig.~\ref{fig:Fig2}, we show the analytical $\dw$ for a string resonator as a blue line, see ``SiN string" entry in Table~\ref{tab:fig_merit} for details~\cite{Ghadimi_2018}. We consider a sample of $N=10^6$ nuclear $^1$H spins (gyromagnetic ratio $\gamma=2\pi\times\SI{42.58}{\mega\hertz/\tesla}$). The gradient of the magnetic field is calculated from a magnetic tip simulation~\cite{Supplement}, and we use $T_2=\SI{100}{\micro\second}$~\cite{Bloch_1946}. In this configuration, simulations indicate a maximum frequency shift of approximately $|\delta\omega_{\text{max}}|/2\pi=\SI{5}{mHz}$ at a detuning of $(\wl-\omega_0)/2\pi\approx \SI{10}{kHz}$ at $T=\SI{4}{K}$. The analytical frequency shift increases twenty-fold to $|\delta\omega_{\text{max}}|/2\pi\approx\SI{100}{mHz}$ when the temperature is lowered to $T=\SI{0.2}{K}$ because of the higher Boltzmann polarization.
 To substantiate our analytical result in Eq.~\eqref{eq:dyn_back_simple}, we perform a numerical simulation of  Eqs.~\eqref{eq:eom_q_fin}-\eqref{eq:eom_Iz_fin}. The simulation closely aligns with the expected results but overestimates the value of $|\delta\omega_{\text{max}}|$ at $\SI{4}{\kelvin}$, as seen in Fig.~\ref{fig:Fig2}(a). The excellent agreement at a lower temperature in Fig.~\ref{fig:Fig2}(b) suggests that deviations arise from the strain on the assumption $\Omega_R \ll \gp, \kappa$ for fixed spin lifetime $T_1$, as $\Omega_R\propto z_0$ increases with thermal motion~\cite{Supplement}.

\begin{figure}[b]
\centering\includegraphics[width=0.48\textwidth]{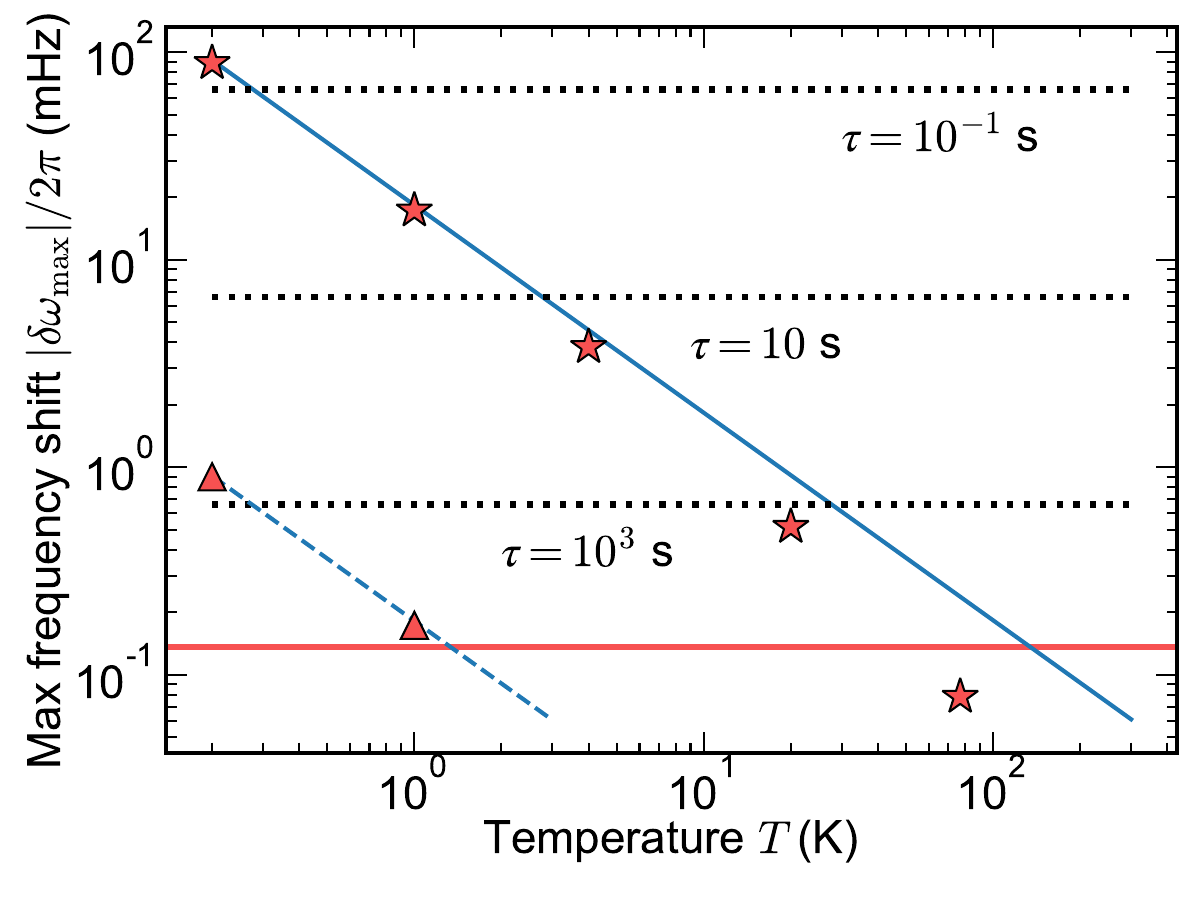}
    \caption{Absolute maximal frequency shift $|\delta\omega_\mathrm{max}|$ vs temperature for the string resonator reported in Table~\ref{tab:fig_merit} (red stars)~\cite{Ghadimi_2018}. 
    Predictions for $10^4$ spins are shown as red triangles. The standard deviation of the thermal frequency noise is displayed for different integration times $\tau$ as black dashed lines, setting the minimum threshold for frequency estimation. The blue lines show the analytical frequency shift for $10^6$ spins (solid) and $10^4$ spins (dashed). The red line indicates the signal that would arise from a single spin. The required averaging time to resolve it is $\tau\sim2.5\times10^4$ s.}
    \label{fig:fig_3}
\end{figure}

\begin{table*}[t]
\caption{Reference resonator parameters. Parameters are estimated or extrapolated from indicated references. Common parameters are $N=10^6$ spins, $T_2 = \SI{100}{\micro\second}$, $T_1 = \SI{50}{\milli\second}$ and resonant driving of the resonator $\wdd=\wo$.} \label{tab:fig_merit}
\centering
\begin{tabularx}{\linewidth}{p{5.cm} *{4}{>{\centering\arraybackslash}X}}
\toprule
Parameters & Symbol & SiN membrane~\cite{Tsaturyan_2017} & SiN string~\cite{Ghadimi_2018} & graphene sheet~\cite{Weber_2016} \\
\midrule
Effective mass (kg) & $m$ & $5\times10^{-12}$ & $2\times 10^{-15}$ & $10^{-17}$ \\
Mechanical frequency (MHz) & $\omega_0/2\pi$ & $1.4$ & $5.5$ & $46$ \\
Spring constant (N/m) & $k$ & 387 & 2.4 & 0.8 \\
Quality factor (low temperature) & $Q/10^6$ & $1000$ & $1000$ & $0.2$ \\
Gradients (MT/m) & $G_{x,y}$, $G_z$ & $2$, $1$ & $2$, $1$ & $2$, $1$\\
Driven amplitude (pm) & $z_0$ & 10 & 1 & 2 \\
Thermal motion (pm) & $z_\mathrm{th}$ & 0.08 & 1 & 1.8 \\
Rabi frequency (Hz) & $\Omega_R/2\pi$ & $850$ & $85$ & $170$  \\
Temperature (K) & $T$ & 0.2 & 0.2 & 0.2 \\
\textbf{Maximum  shift (mHz)} & $|\delta\omega_\mathrm{max}|/2\pi$ & $\mathbf{0.022}$ & $\mathbf{90}$ & $\mathbf{15500}$\\
\bottomrule
\end{tabularx}
\end{table*}

 To assess the performance of our spin detection scheme, we calculate numerical frequency shifts for various device geometries (Table~\ref{tab:fig_merit}), showcasing the versatility of the method. The frequency shift notably increases with a decrease in the resonator's mass, for example graphene sheet resonators are expected to achieve significant frequency shifts exceeding $|\delta\omega_{\text{max}}|/2\pi=\SI{15.5}{kHz}$. Note that the $1/\wdd$ dependence of $g^2$ in Eqs.~\eqref{eq:dyn_back_simple} and \eqref{eq:dyn_back_simple_2} is compensated by $I_0\propto \wl$.
 
 We systematically collect in Fig.~\ref{fig:fig_3} the maximum frequency shift as a function of temperature for string resonators, and compare it with the expected thermal frequency noise~\cite{Ghadimi_2018}. We find that the protocol should allow the detection of $10^6$ nuclear spins ($^1$H) at \SI{4}{\kelvin} within less than a minute of averaging under ideal conditions. At moderate dilution refrigeration temperatures of \SI{0.2}{\kelvin}, the same measurement should be feasible within less than \SI{100}{\milli\second}. We should consider two limitations under realistic conditions: first, technical frequency noise stemming from e.g. temperature drifts will increase the frequency fluctuations. Second, inhomogeneous spectral broadening (a distribution in $\wl$) of a spin ensemble can reduce the signal, which we discuss in the supplemental material~\cite{Supplement}.
 

The theory results summarized in Fig.~\ref{fig:fig_3} demonstrate the advances that are feasible in force-detected nanoscale MRI when considering only Boltzmann (thermal) polarization. This polarization is very small for the relevant temperatures and Larmor frequencies; for instance, at \SI{4}{\kelvin} and $\wl/2\pi = \SI{5.5}{\mega\hertz}$ (corresponding to $|B_0| = \SI{130}{\milli\tesla}$), the effective mean ensemble polarization of a sample containing $10^6$ $^1$H spins is equivalent to that of 33 fully polarized $^1$H spins~\cite{Supplement}. While the detection of a single \textit{electron} spin with a silicon cantilever required an averaging time of roughly \SI{4.7e4}{\second} in 2004~\cite{Rugar_2004}, our method offers the sensitivity for detecting a single \textit{nuclear} spin (with a roughly $1500$ times lower magnetic moment) in \SI{2.5e4}{\second}. A horizontal red line in Fig.~\ref{fig:fig_3} is the corresponding frequency shift for the string device considered here.


The sensitivity of MRI at the nanoscale can be boosted by targeting the statistical spin polarization~\cite{Degen_2007}, whose standard deviation surpasses the mean polarization for small ensembles~\cite{Herzog_2014}. From simple considerations, we indeed expect that spin-mediated resonator backaction can be extended to the statistical regime. There, instead of a static frequency shift as in Eq.~\eqref{eq:dyn_back_simple}, fluctuating nuclear polarization will generate a fluctuating resonator frequency that can be accurately monitored using a phase-locked loop. However, a full model for this regime, based on higher momenta analysis of spin and resonator quantum statistical distributions,~\cite{Kubo_1962}, is left for future work.

Our method uses a single drive (e.g. via electrical or optomechanical coupling) acting directly on the resonator to detect nuclear spins. It thereby significantly reduces the experimental overhead compared to typical MRFM experiments, which require a microstrip close by the resonator~\cite{Poggio_2007} to generate periodic spin flipping through radio-frequency pulses~\cite{Degen_2009}. Dynamic nuclear polarization (DNP) can in principle be implemented to further increase the signal-to-noise ratio~\cite{Carver_1953,Carver_1956}, potentially allowing the detection of a single nuclear spin within minutes of averaging time. Near-resonant spin-mechanics coupling even opens the possibility of coherently manipulating nuclear spins through mechanical driving~\cite{Hunger_2011}. For instance, mechanical vibrations can be harnessed to tilt the ensemble spin polarization fully into the $x$-$y$ plane, resulting in a large increase in the measured signal. Moreover, an intriguing possibility arises when swapping the roles of the resonator and the spin ensemble for spin cooling through backaction~\cite{Butler_2011}, akin to cavity cooling in the reversed dissipation regime in cavity optomechanics~\cite{Nunnenkamp_2014,Ohta_2021}. Our simplified study paves the way for delving into the intricacies of local spin dissipation and decoherence~\cite{Kirton_2017} and dipole-dipole interactions~\cite{Dalla_Torre_2016} in particular experimental configurations. It also lays the groundwork for exploring further opportunities of parametric driving~\cite{Kosata_2020} and multimode resonators~\cite{Dobrindt_2010,Burgwal_2020, Burgwal_2023}. With these capabilities, nanoscale MRI will become a versatile platform for nuclear spin quantum sensing and control on the atomic scale.

\section*{Acknowledgments}
We thank Raffi Budakian, Oded Zilberberg, and Jan Košata for fruitful discussions.
This work was supported by the Swiss National Science Foundation (CRSII5\_177198/1) and an ETH Zurich Research Grant (ETH-51 19-2).

%

\clearpage
\onecolumngrid

\renewcommand{\thesection}{S\arabic{section}}   
\renewcommand{\thetable}{S\arabic{table}}   
\renewcommand{\thefigure}{S\arabic{figure}}
\renewcommand{\theequation}{S\arabic{equation}}

\setcounter{equation}{0}
\setcounter{figure}{0}
\setcounter{table}{0}
\setcounter{page}{1}

\begin{center}
	\textbf{\large Supplemental Material: Near-resonant nuclear spin detection with high-frequency mechanical resonators}
\end{center}
\setcounter{secnumdepth}{2} 

\section{Analytical approach}

\subsection{Master equation formalism}

We offer further details on the analytical solution for the spin-mechanical model introduced in the main text. The model features a driven mechanical resonator moving along $z$, influencing an ensemble of $N$ spins. The spins interact also with a spatially-dependent magnetic field. The combined dynamics is described by the Hamiltonian
\begin{equation}\label{eq:Hamilt_gen}
	\mathcal{H} = \frac{\p^2}{2m}+\frac{1}{2}m\wo^2\q^2 - F_0\q\cos(\wdd t) - \hbar\wl\hat{I}_z -\hbar\gamma\q\left(G_x\hat{I}_x + G_y\hat{I}_y + G_z\hat{I}_z\right),
\end{equation}
where $\hbar$ is the reduced Planck constant, $\wo$ ($\wl$) is the mechanical (Larmor) resonance frequency, $\wdd$ is the driving frequency, $G_i$ is the magnetic gradient along $i\in[x,y,z]$, $\gamma$ is the gyromagnetic ratio of a nuclear spin, and $F_0$ is the driving force. Here $\q$ and $\p$ stand for the position and momentum operators for the resonator. The spins are described by the collective spin operators $\hat{I}_i = \sum_{k=1}^N\hat{\sigma}_{i,k}/2$, where $\hat{\sigma}_{i,k}$ are the Pauli matrices describing a spin-$\frac{1}{2}$.

We extract the dissipative equations of motion (EOM) using the Heisenberg picture. We account for mechanical damping ($\Gm$) as well as spin decay ($T_1$) and decoherence ($T_2$). In order to get physically measurable quantities, we take the mean value of the derived EOM, these are given by:
\begin{align}
	\lb\ddot{\q}\rb &= -\wo^2\lb\q\rb -\Gm\lb\dot{q}\rb + \frac{F_0}{m}\cwd + \frac{\hbar\gamma}{m}\left(G_x\lb\hIx\rb+G_y\lb\hIy\rb+G_z\lb\hIz\rb\right), \label{eq:eom_q} \\
	\lb\dot{\hat{I}}_{x,y}\rb &=-\frac{1}{T_2}\lb \hat{I}_{x,y}\rb \pm \wl\lb \hat{I}_{y,x}\rb \pm\gamma G_z\lb\q \hat{I}_{y,x}\rb \mp \gamma G_{y,x}\lb\q\hIz\rb, \label{eq:eom_Ixy} \\
	\lb\dhIz\rb &= \frac{1}{T_1}\left(I_0 - \lb\hIz\rb\right) - \gamma G_x\lb\q\hIy\rb + \gamma G_y\lb\q\hIx\rb, \label{eq:eom_Iz}
\end{align}
where we identify the renormalized mechanical frequency as $\wo\mapsto\sqrt{\omega_0^2+\Gamma_m^2/4}$. Here $I_0$ stands for the Boltzmann (thermal) equilibrium polarization~\cite{Niinikoski_2020} 
\begin{align}\label{eq:Boltzmann_pol}
	I_0=-N\left[(2I+1)\coth\left((2I+1)\hbar\wl/(2k_BT)\right)-\coth(\hbar\wl/(2k_BT))\right]/2,
\end{align}
with $N$ the number of spins in the considered ensemble and $I=\frac{1}{2}$ the spin number.

The resonator motion is driven well above its zero-point fluctuation, we can therfore apply the mean-field approximation and study the semiclassical limit. This entails separating the cross-correlated mean values, i.e. $\lb\q\hat{I}_i\rb = \lb\q\rb\lb\hat{I}_i\rb$. We furthermore drop the braket and hat notation, e.g. $\lb\hIx\rb \equiv I_x$, leading to the EOM presented in the main text Eqs.~(4)-(6), namely,
\begin{align}
	\ddot{q} &=  -\wo^2q -\Gm\dot{q} + \frac{F_0}{m}\cwd + \frac{\hbar\gamma}{m}\textbf{G}\cdot \textbf{I},\label{eq:eom_q_fin_S}\\
	\dot{I}_{x,y}&=-\frac{1}{T_2}I_{x,y}\pm(\wl+\gamma qG_{z})I_{y,x}\mp\gamma qG_{y,x}I_{z}\label{eq:eom_Ixy_fin_S} ,\\
	\dIz &= \frac{1}{T_1}\left(I_0 - \Iz\right) - \gamma q\left(G_x\Iy - G_y\Ix\right),\label{eq:eom_Iz_fin_S}
\end{align}
with vectors $\textbf{G}=(G_x,G_y,G_z)$ and $\textbf{I}=(I_x,I_y,I_z)$.

\subsection{Slow-flow equations of motion}
We further analyze here the solution to the main text Eqs. (4)-(6). We assume that the resonator dynamic is dominated by the external force. Thus, the coupling to the spins act as a small correction. We can then write the mechanical motion as $q(t)=q^{(0)}(t)+\delta q(t)$, where $q^{(0)}(t)=u_q^{(0)}\cwd+v_q^{(0)}\swd$ with 
\begin{align}
	u_q^{(0)} =& \frac{F_0}{m\left[\left(\wo^2-\wdd^2\right)^2 + \wdd^2\Gm^2\right]}\left(\wo^2-\wdd^2\right),\\
	v_q^{(0)} =& \frac{F_0}{m\left[\left(\wo^2-\wdd^2\right)^2 + \wdd^2\Gm^2\right]}\wdd\Gamma_m.
\end{align}
The part accounting for the spins then obeys
\begin{equation}
	\ddot{\delta q}=-\omega_{0}^{2}\delta q-\Gamma_{m}\dot{\delta q}+\frac{\hbar\gamma}{m}\textbf{G}\cdot\textbf{I}.
\end{equation}

We express the solution for $\delta q(t)$ in terms of an ansatz
\begin{equation}\label{eq:sol_driven_oscill_delta}
	\delta q(t) = \delta a_q(t) + \delta u_q(t)\cwd + \delta v_q(t)\swd,
\end{equation}
where $\delta a_q(t)$, $\delta u_q(t)$ and $\delta v_q(t)$ are real time-dependent amplitudes to be found. Employing this form of the solution is particularly beneficial when examining perturbations associated with the behavior of a driven harmonic oscillator.   The dynamics of the spins in response to the mechanical motion can be calculated employing a similar ansatz
\begin{equation}\label{eq:ansatz_spins}
	I_i(t) = a_i(t) + u_i(t)\cwd + v_i(t)\swd,
\end{equation}
with amplitudes $a_i(t), u_i(t), v_i(t)$. Given Eq.~\eqref{eq:ansatz_spins}, the spins exert a time-dependent force on the resonator given by 
\begin{align}\label{eq:spin_force}
	\delta F(t)=\hbar\gamma\textbf{G}\cdot\textbf{I}(t)=\hbar\gamma\left[\mathbf{G}\cdot\mathbf{a}(t)+\mathbf{G}\cdot\mathbf{u}(t)\cos(\omega_{d}t)+\mathbf{G}\cdot\mathbf{v}(t)\sin(\omega_{d}t)\right], 
\end{align}
where we used vector notation for $a_{x,y,z}(t), u_{x,y,z}(t), v_{x,y,z}(t)$.  

At this stage, we have not yet introduced any constraints or approximations in the ansatz amplitudes.  However, the calculation of the corrections $\delta a_q,\delta u_q,\delta v_q$ is greatly facilitated by assuming the weak impact of the spin-dependent force on the resonator. Namely, we assume $\langle\langle|\hbar\gamma\textbf{G}\cdot \textbf{I}|\rangle\rangle_{T_d}\ll F_0$, where $\langle\langle...\rangle\rangle_{T_d}$ denotes the average over a drive period $T_{\mathrm{d}}=2\pi/\wdd$.  In this setting, we can assume the amplitudes  $\delta a_q(t),\delta u_q(t),\delta v_q(t)$ with respect to $T_{\mathrm{d}}$, accounting for the transient evolution of the amplitude and phase of the resonator towards the steady state~\cite{Rand_2005}. 

In the steady state, resonator and spin precession amplitudes settle to constant values, i.e. $\delta\dot{a}_q=\delta \dot{u}_q=\delta \dot{v}_q=0$ and $\dot{a}_i=\dot{u}_i=\dot{v}_i=0$. In our ansatz $q(t)$ thus acts as a harmonic magnetic field with frequency $\wdd$, acting on the spins. In particular, the spin prompts a spin precession component at frequency $\wdd$, according to Eq.~\eqref{eq:ansatz_spins}. Note the ansatz does not presuppose the synchronization or ``locking'' of the spin dynamics with the external field. We seek if such a steady state can exist.
To this end, we insert the ansatz for $q(t)$ and $I_i(t)$ in the mean-field equations of motion and equate the harmonic amplitudes at both sides of the equations with the same time dependence, a procedure dubbed the ``harmonic balance''~\cite{Krack_2019}. This approach also neglects super-harmonic generation (e.g. terms $\cos(2\wdd t)$, $\sin(2\wdd t)$) that arises from the mechanical motion driving the spins, which requires extending the harmonic ansatz for $q(t), I_i(t)$ to higher frequencies. Note that harmonic balance relies on the slowly-flowing nature of the amplitudes $a_i, u_i, v_i$~\cite{Rand_2005}.

\subsubsection{Linear response theory}

The introduction of the ansatz results in nonlinear couplings between the harmonic amplitudes of the mechanical resonator and the spins. The system's steady states are defined by the roots of these coupled polynomials. While we could solve these equations numerically using advanced algebraic methods, as detailed in reference~\cite{Breiding_2018} and implemented in the package~\cite{Kosata_2022_SP}, we opt for deriving an analytical solution within a linearized framework.     Here we find the mechanical dynamics of the resonator in the weakly fluctuating regime $\langle\langle|\delta q|\rangle\rangle_{T_d}\ll \langle\langle |q^{(0)}|\rangle\rangle_{T_d}$. The smallness of $\delta q$ allows us to neglect the nonlinear coupling between the fluctuations $\delta u_q(t),\delta v_q(t)$ and the spin amplitudes $a_i(t),u_i(t),v_i(t)$. Under this linearization, the spin dynamics directly follows from the solutions of the first-order differential equations that do not contain $\delta a_q(t),\delta u_q(t),\delta v_q(t)$, namely
\begin{subequations}\label{eq:spin_eqs_linear}
	\begin{align}
		\dot{a}_{x,y} + \gp a_{x,y} \mp \wl a_{y,x} &= 0, \\
		\dot{u}_{x,y} + \gp u_{x,y} \mp \wl u_{y,x} + \wdd v_{x,y} - \gamma u_q^{(0)}(G_{z,x}a_{y,z} - G_{y,z}a_{z,x}) &= 0, \\
		\dot{v}_{x,y} + \gp v_{x,y} \mp \wl v_{y,x} - \wdd u_{x,y} - \gamma v_q^{(0)}(G_{z,x}a_{y,z} - G_{y,z}a_{z,x}) &= 0,\\ 
		\dot{a}_z + \frac{1}{T_1} a_z - I_0\frac{1}{T_1} &= 0, \\
		\dot{u}_z + \frac{1}{T_1} u_z + \wdd v_z - \gamma u_q^{(0)}(G_ya_x - G_xa_y) &= 0, \\
		\dot{v}_z + \frac{1}{T_1} v_z - \wdd u_z - \gamma v_q^{(0)}(G_ya_x - G_xa_y) &= 0.
	\end{align}
\end{subequations}

The resonator features a high quality factor ($\Gamma_m\ll\gp,\frac{1}{T_1}$) which, together with the weak spin-resonator coupling ($\gamma G_i z_0\ll \gp,\frac{1}{T_1}$) lead to spins quickly reaching steady state compared to the slower resonator timescale. This condition permits the application of approximation methods, like adiabatic elimination of the spins~\cite{Lugiato_1984}, in order to approximate the time evolution of the resonator towards its steady state. Our focus is nevertheless on the global steady state behavior, where all amplitudes in the problem are fixed.  Solving Eqs.~\eqref{eq:spin_eqs_linear} when $\dot{a}_i=\dot{u}_i=\dot{v}_i=0$ to find the steady state amplitudes $\textbf{a}|_{t\rightarrow\infty},\textbf{u}|_{t\rightarrow\infty},\textbf{v}|_{t\rightarrow\infty}$ leads to a steady state force
\begin{equation}\label{eq:steady_force}
	\delta F|_{t\rightarrow\infty}\approx\hbar\gamma\left[\mathbf{G}\cdot\mathbf{a}|_{t\rightarrow\infty}+\mathbf{G}\cdot\mathbf{u}_{t\rightarrow\infty}\cos(\omega_{d}t)+\mathbf{G}\cdot\mathbf{v}_{t\rightarrow\infty}\sin(\omega_{d}t)\right].
\end{equation}

Such force \eqref{eq:steady_force} will not be in phase with the external resonator's driving (its quadratures will not be aligned with the drive), namely $u_q^{(0)}$, $v_q^{(0)}$. To facilitate the expressions, we choose a phase/time origin for the driven resonator (i.e. we perform a gauge fixing), such that $v_q^{(0)}=0$ and $u_q^{(0)}=F_0/(m\sqrt{\left(\wo^2-\wdd^2\right)^2 + \wdd^2\Gm^2})$. In this gauge, we can identify $\delta F|_{t\rightarrow\infty}\approx \delta F_0 - \delta\Gm\dot{q} - \delta\Omega^2 q$, where $\delta F_0 = \hbar\gamma I_0G_z/m$ and
\begin{align}
	\delta\Gm &= \frac{\hbar\gamma^2I_0\left(G_x^2+G_y^2\right)}{m}\frac{\wl T_2^{-1}}{\left(T_2^{-2} + \wdd^2\right)^2 + 2\left(T_2^{-1}-\wdd\right)\left(T_2^{-1}+\wdd\right)\wl^2 +\wl^4}, \\
	\delta\Omega^2 &= -\frac{\hbar\gamma^2I_0\left(G_x^2+G_y^2\right)}{m}\frac{\wl\left(T_2^{-2}-\wdd^2+\wl^2\right)}{\left(T_2^{-2} + \wdd^2\right)^2 + 2\left(T_2^{-1}-\wdd\right)\left(T_2^{-1}+\wdd\right)\wl^2 +\wl^4}. 
\end{align}
We can now reconstruct the mechanical evolution in the steady state from the effective equation of motion. Under resonant driving  $\wdd=\wo$,
\begin{equation}
	\ddot{q} + \left(\Gm+\delta\Gm\right)\dot{q} + \left(\wo + \dw\right)^2q |_{t\rightarrow\infty} = F_0\cwd + \delta F_0, \label{eq:sol_memb_harmonic}
\end{equation}
with $\dw = \frac{1}{2}\frac{\delta\Omega^2}{\wdd}$. We can rewrite $\delta\Gm$ and $\dw$ in a more convenient way leading to main text Eqs. (7) and (8)
\begin{align}
	\dw = -g^2\left(\frac{\wl + \wdd}{1/T_2^2 + \left(\wl+\wdd\right)^2} + \frac{\wl - \wdd}{1/T_2^2+\left(\wl-\wdd\right)^2}\right), \\
	\delta\Gm = -g^2\left(\frac{T_2^{-1}}{1/T_2+\left(\wl+\wdd\right)^2} - \frac{T_2^{-1}}{1/T_2^2 + \left(\wl-\wdd\right)^2}\right),
\end{align}
where we introduced $g^2 = \hbar\gamma^2I_0\left(G_x^2+G_y^2\right)/(4m\wdd)$.

\subsubsection{Beyond linear response}

\begin{figure}[t]
	\centering
	\includegraphics[width=0.6\textwidth]{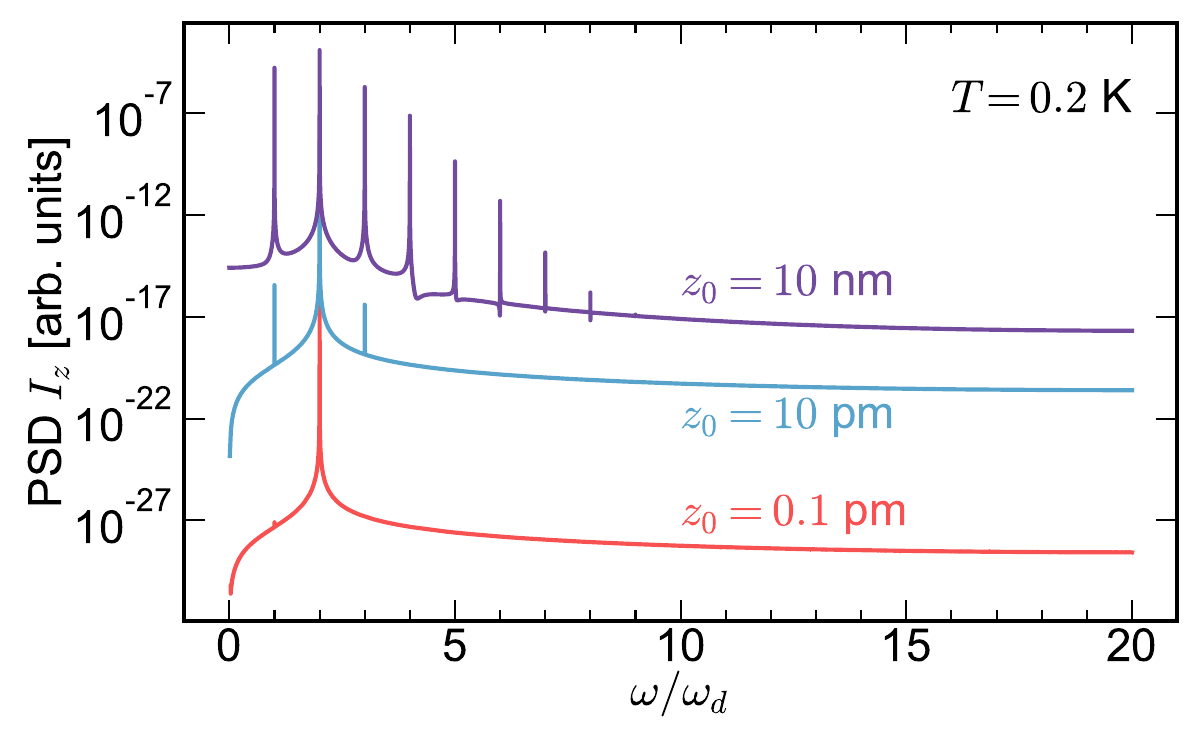}
	\caption{Power spectral density (PSD) of the steady state of a simulated SiN membrane resonator with zero frequency component removed~\cite{Tsaturyan_2017} for different driving amplitudes $z_0$, $m=\SI{5e-12}{\kilo\gram}$, $\wo/2\pi=\SI{1.4}{\mega\hertz}$, $G_x=G_y=\SI{2}{\mega\tesla/\meter}$, $G_z=\SI{1}{\mega\tesla/\meter}$, $N=10^6$ spins, $T_2 = \SI{100}{\micro\second}$, $T_1=\SI{50}{\milli\second}$ and resonant driving of the resonator $\wdd=\wo$.}
	\label{fig:higher_orders}
\end{figure}

For certain parameter regimes, the nonlinearities in Eqs.~\eqref{eq:eom_q_fin_S}-\eqref{eq:eom_Iz_fin_S} can lead to complex behavior in the stationary limit $t\rightarrow\infty$, including self-sustained motion, multi-stability, and limit cycles~\cite{Bhaseen_2012, Chitra_2015}. In particular, the analogy with optomechanics is expected to break down when the Rabi frequency is comparable to the spin dissipation: $\gamma G_iz_0 \sim \gp, \frac{1}{T_1}$. In that case, the spins' equilibration is not fast enough before they act back on the resonator, and spin-resonator timescales cannot be adiabatically separated. Effectively, the resonator motion then triggers spin-induced nonlinear effects, such as a  periodic time modulation of the Larmor frequency due to the $G_z$ gradient (see main text Eqs.~(4)-(6)), with frequency $\wdd$. The resonator's response would then pick up higher frequency components not described by Eq.~\eqref{eq:ansatz_spins}. While under the linearized theory, the steady state value is time independent and equal to $I_z=I_0$, we observe the generation of higher order harmonics in the spectrum of $I_z$ [Fig.~\ref{fig:higher_orders}]. Note that in our simulations, we do not focus on the regime where higher excitation makes the spin-conservation constraint ($d(\sum_i I_i^2)/dt = 0$) relevant, which would lead to additional `many-wave mixing'.

A comprehensive examination of nonlinearities in our detection protocol is beyond the scope of this manuscript; however, we offer a brief overview of the necessary approach below. The impact of weak nonlinearities can be expressed by still expanding the solution for $x(t)$ with an ansatz of the form
\begin{equation}
	x(t)=a_q(t)+u_q(t)\cos(\wdd t) + v_q(t)\sin(\wdd t),
\end{equation}
which includes both the displacement by the driving field and small fluctuations, while keeping the same ansatz for the spins in Eq.~\eqref{eq:ansatz_spins}. We will account now for the nonlinear corrections to this resonant behavior. The equations of motion for the ansatz amplitudes  without linearization read 
\begin{align}\label{eq:nonlinear_evo}
	&\omega_0^{2} a_q   + \tilde{G}_x a_x + \tilde{G}_y a_y + \tilde{G}_z a_z+ \Gamma_m\dot{a}_q=0, \\
	&\omega_0^{2} u_q - \wdd^2 u_q + \Gamma_m\wdd v_q + 2 \wdd \dot{v}_q  -F_0 + \tilde{G}_x u_x + \tilde{G}_y u_y + \tilde{G}_z u_z +\Gamma_m\dot{u}_q = 0 , \\
	&\omega_0^{2} v_q  - \wdd^2 v_q - 2 \wdd \dot{u}_q - \Gamma_m\wdd u_q   \tilde{G}_x v_x + \tilde{G}_y v_y + \tilde{G}_z v_z + \Gamma_m\dot{v}_q= 0 , \\
	&\frac{1}{T_2} a_x + \wl a_y + \tilde{G}_z a_q a_y - \tilde{G}_y a_q a_z - \frac{\tilde{G_y}}{2} u_q u_z + \frac{\tilde{G_z}}{2} u_q u_y + \frac{\tilde{G_z}}{2} v_q v_y - \frac{\tilde{G_y}}{2} v_q v_z + \dot{a}_x = 0, \\
	&\frac{1}{T_2} u_x + \wl u_y + \wdd v_x + \tilde{G}_z a_q u_y + \tilde{G}_z a_y u_q - \tilde{G}_y a_q u_z - \tilde{G}_y a_z u_q + \dot{u}_x = 0, \\
	&\frac{1}{T_2} v_x + \wl v_y - \wdd u_x + \tilde{G}_z a_q v_y + \tilde{G}_z a_y v_q - \tilde{G}_y a_z v_q - \tilde{G}_y a_q v_z + \dot{v}_x = 0, \\
	&\frac{1}{T_2} a_y - \wl a_x + \tilde{G}_x a_q a_z - \tilde{G}_z a_q a_x - \frac{\tilde{G_z}}{2} u_q u_x + \frac{\tilde{G_x}}{2} u_q u_z + \frac{\tilde{G_x}}{2} v_q v_z - \frac{\tilde{G_z}}{2} v_q v_x + \dot{a}_y = 0, \\
	&\frac{1}{T_2} u_y + \wdd v_y - \wl u_x + \tilde{G}_x a_q u_z + \tilde{G}_x a_z u_q - \tilde{G}_z a_q u_x - \tilde{G}_z a_x u_q + \dot{u}_y = 0, \\
	&\frac{1}{T_2} v_y - \wdd u_y + \tilde{G}_x a_z v_q - \wl v_x + \tilde{G}_x a_q v_z - \tilde{G}_z a_x v_q - \tilde{G}_z a_q v_x + \dot{v}_y = 0, \\
	&\frac{1}{T_1} a_z - I_0 \frac{1}{T_1} + \tilde{G}_y a_q a_x - \tilde{G}_x a_q a_y + \frac{\tilde{G_y}}{2} u_q u_x - \frac{\tilde{G_x}}{2} u_q u_y - \frac{\tilde{G_x}}{2} v_q v_y + \frac{\tilde{G_y}}{2} v_q v_x + \dot{a}_z = 0, \\
	&\frac{1}{T_1} u_z + \wdd v_z + \tilde{G}_y a_q u_x + \tilde{G}_y a_x u_q - \tilde{G}_x a_y u_q - \tilde{G}_x a_q u_y + \dot{u}_z = 0, \\
	&\frac{1}{T_1} v_z - \wdd u_z + \tilde{G}_y a_q v_x + \tilde{G}_y a_x v_q - \tilde{G}_x a_q v_y - \tilde{G}_x a_y v_q + \dot{v}_z = 0.
\end{align}

where $\tilde{G}_i=\gamma G_i$ is shorthand.  The resonator's susceptibility can then by found by (i) finding the steady states of these equations, i.e. finding the roots of a system of coupled polynomials arising from $\dot{a}_i=\dot{u}_i=\dot{v}_i=\dot{a}_q=\dot{u}_q=\dot{v}_q=0$, and (ii) performing
linear fluctuation analysis around these solutions. These two steps can be facilitated by the use of the HarmonicBalance.jl package~\cite{Kosata_2022_SP}.

Finally, the frequency spectrum in Fig.~\ref{fig:higher_orders} reveals that as the driving strength increases, the lowest order nonlinear effect is the generation of a second harmonic at a frequency $2\wdd$. Similar equations to Eqs.~\eqref{eq:nonlinear_evo} can be similarly obtained for the amplitudes of an extended ansatz that includes also the higher harmonic generated at $2\wdd$.

\subsection{Linewidth change}

\begin{figure}[H]
		\subfloat{\includegraphics[width=0.48\textwidth]{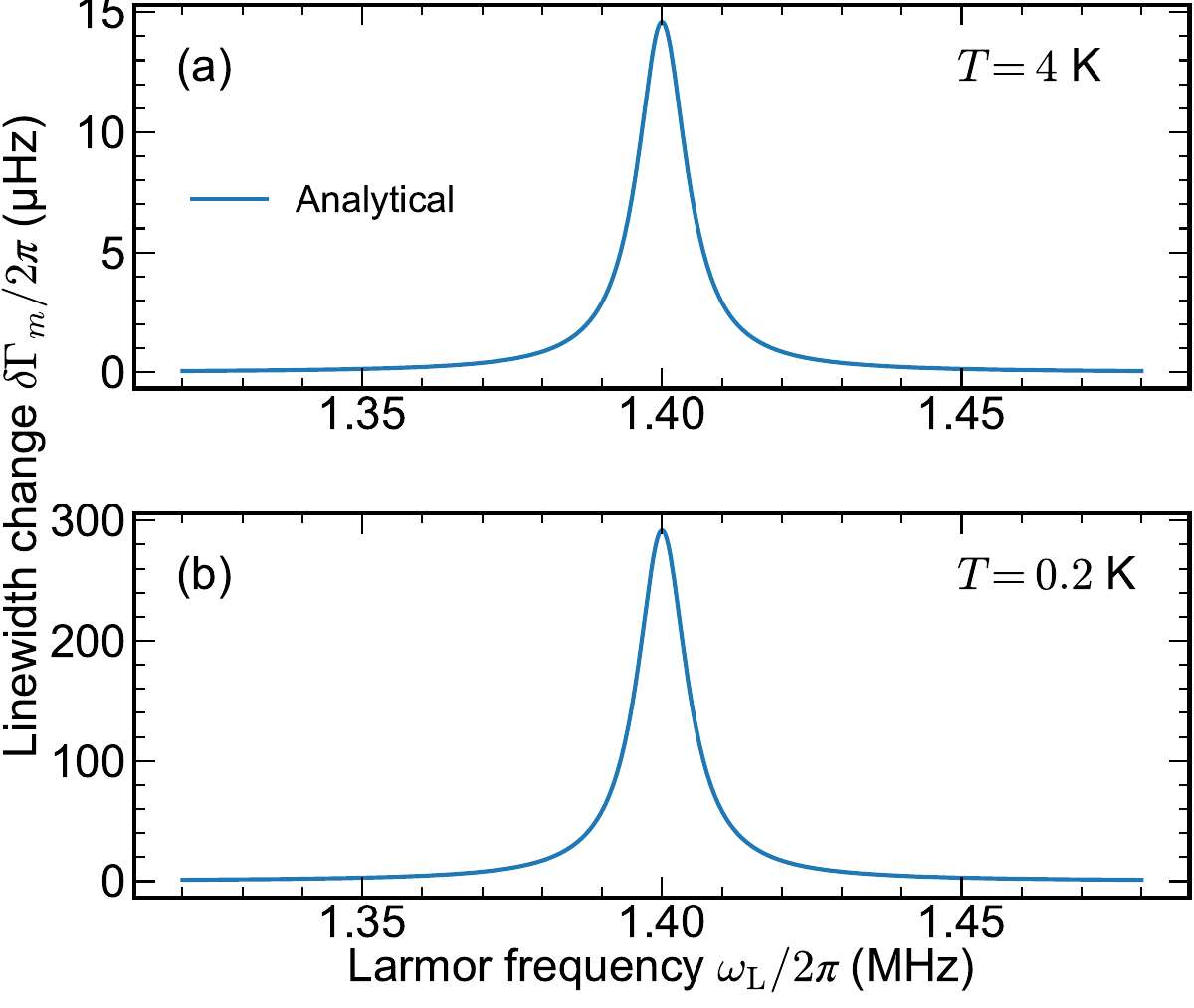}
		\label{fig:linewidth_memb}}
	\hspace{\fill} 
		\subfloat{\includegraphics[width=0.48\textwidth]{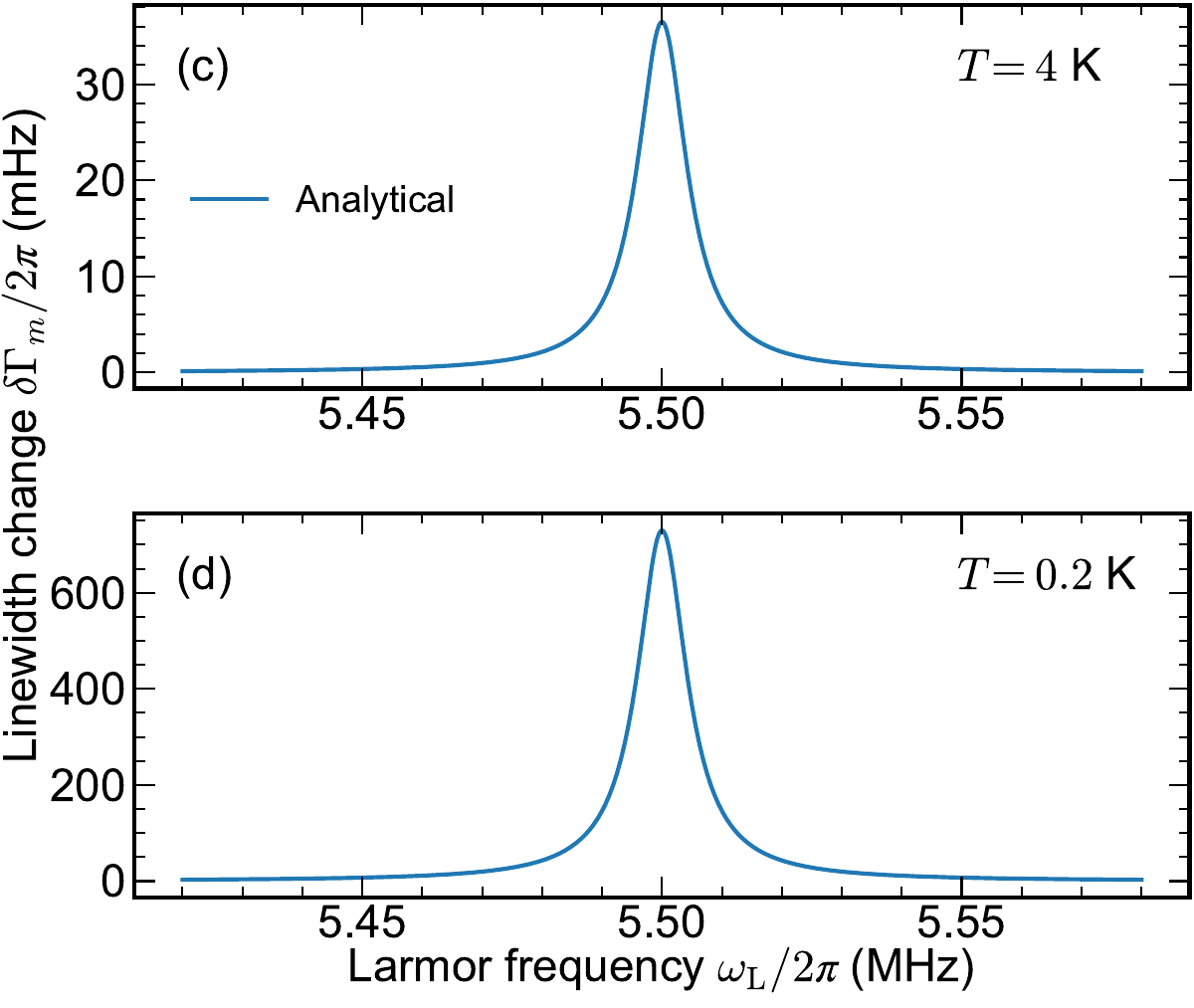}
		\label{fig:linewidth_string}}
	
		\hspace{0.25\columnwidth}
				\subfloat{\includegraphics[width=0.48\textwidth]{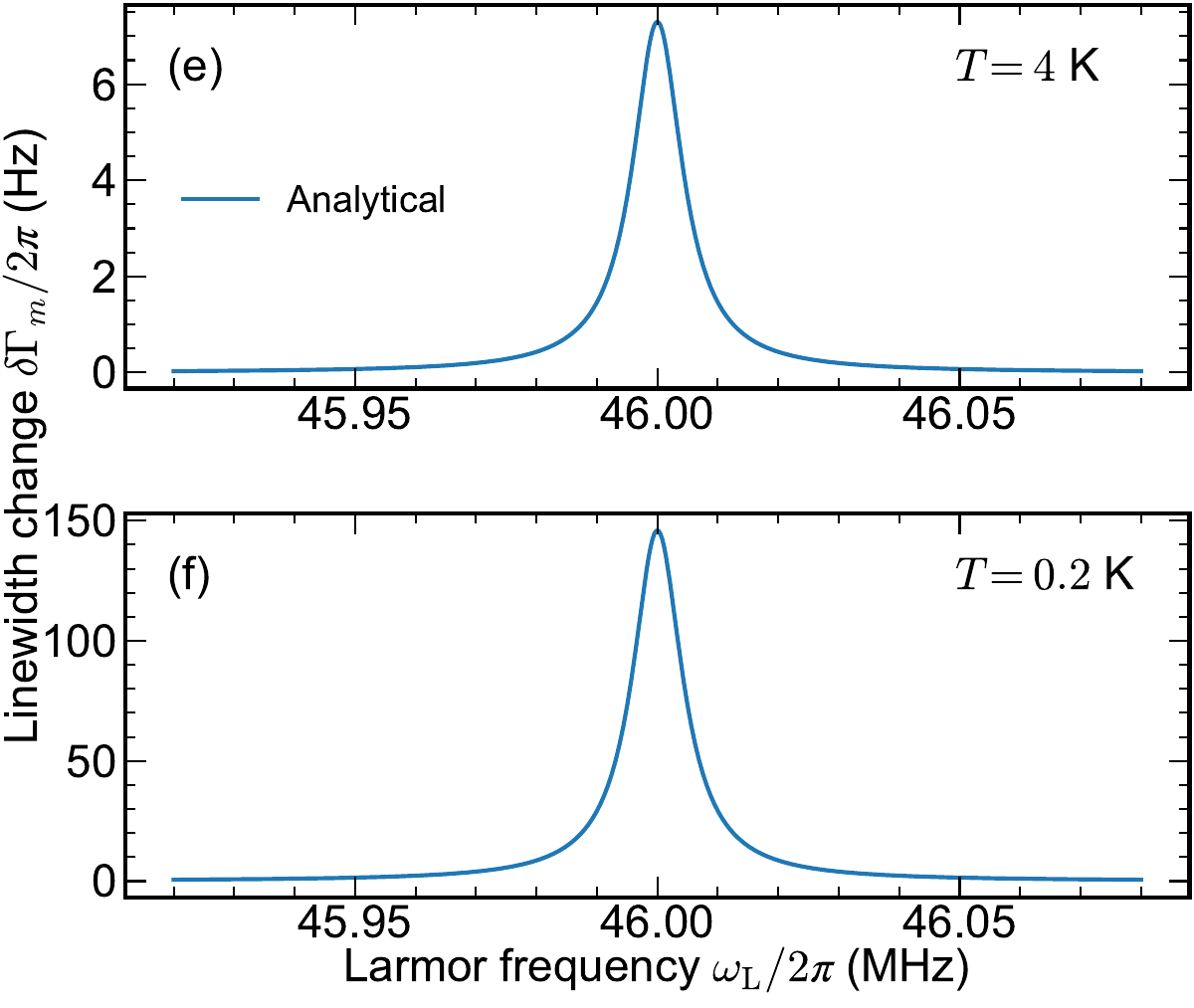}
			\label{fig:linewidth_graphene}}
	\caption{Linewidth change for (a)-(b) a SiN membrane resonator~\cite{Tsaturyan_2017}, (c)-(d) a SiN string resonator~\cite{Ghadimi_2018} and (e)-(f) a graphene sheet resonator~\cite{Weber_2016}, note the difference in the vertical scale unit.}
	\label{fig:linewidth}
\end{figure}

In Fig.~\ref{fig:linewidth}, we show the linewidth change predicted by the analytical model for the three resonators considered in this work. This change produces an effective reduction of the resonator's quality factor, resulting in ``cold damping'', as conventionally used in cavity optomechanics~\cite{Aspelmeyer_2014}. The influence of cold damping on a cantilever in an MRFM experiment, albeit in a nonresonant setting, was studied previously~\cite{Greenberg_2009}.

\section{Relaxation and spectral broadening}

\subsection{Relaxation rates}
The spin lifetime $T_1$ of nuclear spins resulting from energy relaxation can vary strongly in typical nuclear magnetic resonance (NMR) experiments, ranging from microseconds to days. Our experimental situation is untypical, as we will probe nanoscale samples at low temperatures and low magnetic fields. We do not need a very specific value for $T_1$, as our analytical results hold as long as $T_1\ll 1/\Gm$. To avoid speculation about the dependency of $T_1$ on field strength and temperatures below \SI{70}{\kelvin}, we use the same value of $T_1 = \SI{50}{\milli\second}$ for all our simulations. If needed for specific experimental situation, we envisage reducing $T_1$ by introducing paramagnetic agents, such as free radicals or metal ions~\cite{Kocman_2019}.

\subsection{Decoherence due to spin-spin coupling}
In typical NMR experiments, interactions between neighboring nuclear spins can often be neglected when the Rabi frequency $\Omega_R$ exceeds the spin-spin coupling strength $J$. In that scenario, the range of Larmor frequencies that are affected by the spin lock is dominated by the spectral ``power broadening'' equal to $\Omega_R$. By contrast, in the experiments we describe, the condition $\Omega_R\geq J$ is typically not fulfilled, and we are limited by the condition $\Omega_R \ll \frac{1}{T_2}, \frac{1}{T_1}$. As we cannot ignore spin-spin interaction in the weak-driving regime, $J$ is explicitly implemented in the simulations through the spin decoherence time $T_2 = \SI{100}{\micro\second}$~\cite{Bloch_1946}.

\subsection{Inhomogeneous broadening}    
In our paper, we primarily consider the interaction between a mechanical resonator and an ensemble of spins with identical Larmor frequencies. This was done to present the fundamental concept as clearly as possible. However, any real sample is extended in space, and therefore contains spins at various positions within the magnetic field gradient. When all these spins in a sample are excited at the same time, their variation in Larmor frequencies leads to a so-called ``inhomogeneous broadening'' $1/T_2^*$ and a corresponding dephasing time $T_2^*$. For $T_2^*<T_2$, the inhomogeneous broadening dominates over spin-spin interactions and increases the effective spectral with of the ensemble.
In our sample, the driving fields are necessarily weak to fulfill the condition $\Omega_R = \gamma G_i z_0 \ll 1/T_2, 1/T_1$. Only the spins within the narrow range $\wl = \omega_0 \pm \Omega_R$ are directly excited, yielding an inhomogeneous broadening of $1/T_2^* = \Omega_R$. Through spin-spin interactions, all the spins within $\wl = \omega_0 \pm 1/T_2$ are indirectly excited. As $\Omega_R \ll 1/T_2$, the broadening of the spin ensembles is limited by $1/T_2$, not $1/T_2^*$. We therefore do not need to be concerned about the effects of inhomogeneous broadening.

\subsection{Signal contributions with different signs}

The spectral width of the spins that are excited in our scheme depends on $1/T_2$. In principle, this has the consequence that the resonator interacts with spins covering a significant part of the frequencies shown in Fig.~2, which can lead to a mutual cancellation of positive and negative frequency shifts. In the following, we discuss how this problem can be avoided.

\begin{itemize}
	\item \textbf{Differential measurements:} the contributions from different parts of the sample can cancel each other when they generate frequency shifts with equal strength and opposite signs. However, as the nanomagnetic probe is scanned over the sample, spins at different locations enter the strong field gradient at different scan positions. For instance, at a certain position, mainly spins contributing positive frequency shifts may be located within the strong gradient, causing a strong positive frequency shift. The outcome of a full scan then corresponds to a \textit{differential spin signal}, where the frequency shift of the resonator does not indicate the number of spins at one particular Larmor frequency, but rather the relative weight between `positive' and `negative' spins. Knowing the field distribution of the nanomagnetic probe (which can be calibrated with well-known samples), the spatial distribution of spins can be reconstructed by integrating over the differential signal.
	\item \textbf{Statistical polarization:} in MRFM, we typically probe the statistical polarization of a spin ensemble, not the Boltzmann polarization~\cite{Degen_2007}. In this case, only the number of spins (weighted by the square of the local gradient) is important to evaluate the measured signal, not the sign of their individual contributions. For instance, let us assume the simple case of two equal spin sub-ensembles with different Larmor frequencies that contribute with equally strong but opposite frequency shifts in Fig.~2. Assuming the spins fluctuate thermally (and ignoring any remaining Boltzmann polarization), the two ensembles will spend an equal amount of time in a symmetric configuration as in the antisymmetric one. In the antisymmetric configuration, the sub-ensembles will contribute to the frequency shifts with the same sign. The total resonator frequency variance resulting from the combined effect of the two ensembles in the long-time limit is therefore overall \textit{increased}, not reduced. 
	\item \textbf{Single spin:} One exciting aspect of the method we propose is the quantum limit of probing a single spin. In this situation, the problem arising from variations in the Larmor frequency naturally vanishes, leaving only a well-defined Larmor frequency generating a single frequency shift.
\end{itemize}

\section{Exact numerical simulations}
To verify our theoretical predictions, we numerically simulate our mean-field EOM given by Eqs. \eqref{eq:eom_q_fin_S}-\eqref{eq:eom_Iz_fin_S}. We solve the EOM using an explicit Runge-Kutta method of order 8, which is well-suited for handling the large separation of timescales in the problem~\cite{Hairer_1993}. 

\subsection{Results for different devices}
We focus on three device families: silicon nitride (SiN) membranes, SiN strings, and graphene sheets. These devices span across a large range of frequencies and across several orders of magnitude of mass. We simulate a hypothetical sample containing $N=10^6$ $^1$H nuclear spins, corresponding  to a sample size of $\approx(\SI{20}{\nano\meter})^3$~\cite{Krass_2022}. To estimate the magnetic field gradients, we numerically simulate a cobalt nanorod of length \SI{1}{\micro\meter} and radius \SI{50}{\nano\meter} that is pre-magnetized to $\SI{1}{\tesla}$, inspired by Ref.~\cite{Longenecker_2012}. The magnetic simulation results are presented in Sect.~\ref{sub_sect:magnetic_sim}. The driving strength,  proportional to $z_0$, can be easily tuned in the experiment. We choose it in the following such that $\Omega_R\ll \frac{1}{T_2},\frac{1}{T_1}$. Due to the low $T_2=\SI{100}{\micro\second}$ value, this condition might be hard to fulfill for high temperatures as the driving amplitude $z_0$ is lower bound by the resonator's thermal motion. Membrane and string resonators can have quality factor around $Q=\wo/\Gm =10^8$ at room temperature and increase up to $10^9$ for low temperature $T\leq4$ K. To fulfill the condition $\Gm\ll\gp,\frac{1}{T_1}$ with these $Q$ factors, $T_1$ should be at most in the minute range. However, in order to reduce the number of timesteps required in the simulation, we use a reduced $T_1$ time as well as a reduced quality factor $Q$ for the resonator within the regime $\Gm \ll \gp,\frac{1}{T_1}$. Throughout the simulations, we used $T_1=\SI{50}{\milli\second}$.

\begin{figure}[H]
		\subfloat{\includegraphics[width=0.48\textwidth]{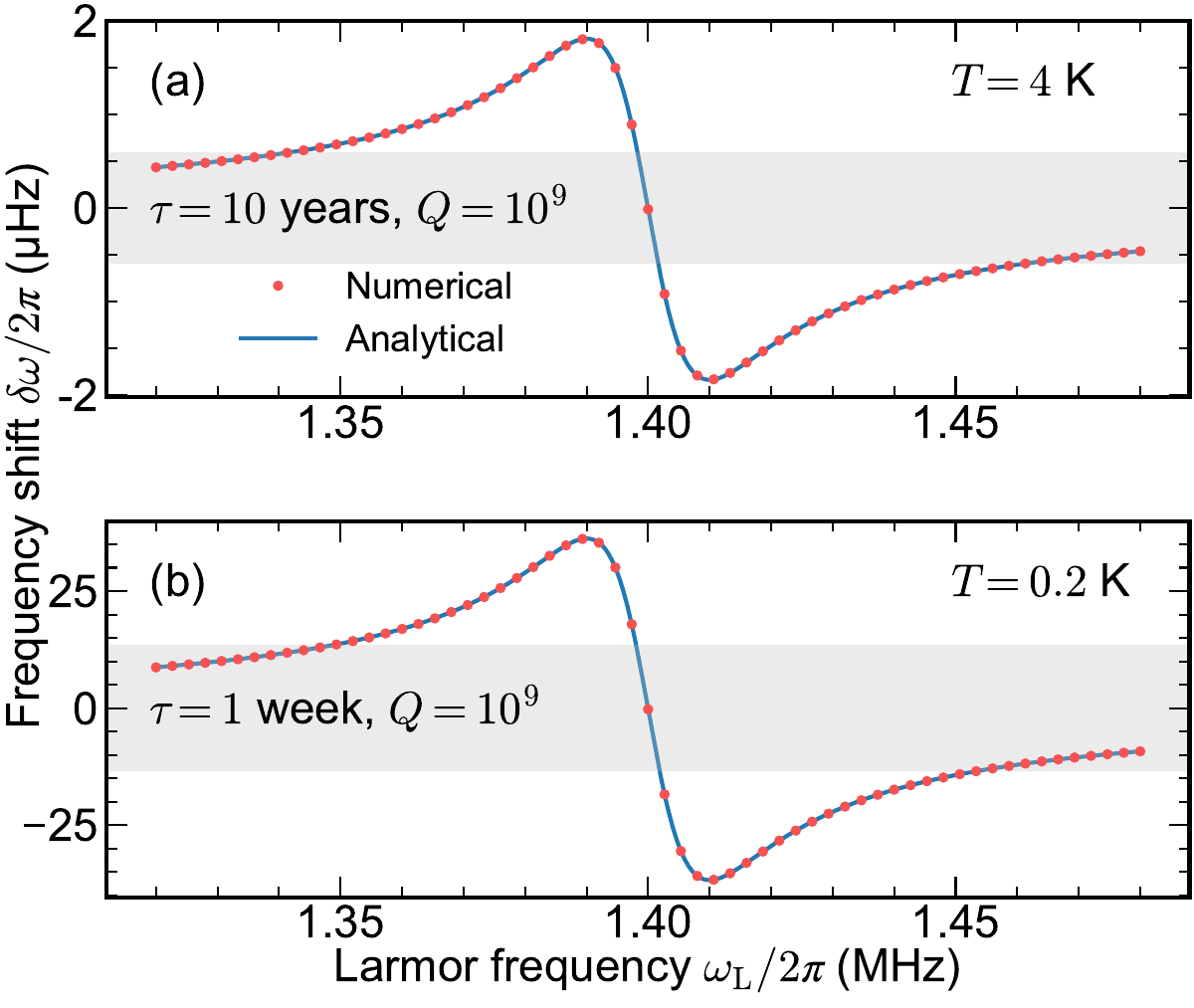}
		\label{fig:membrane_analy_thermal}}
	\hspace{\fill} 
	\subfloat{\includegraphics[width=0.48\textwidth]{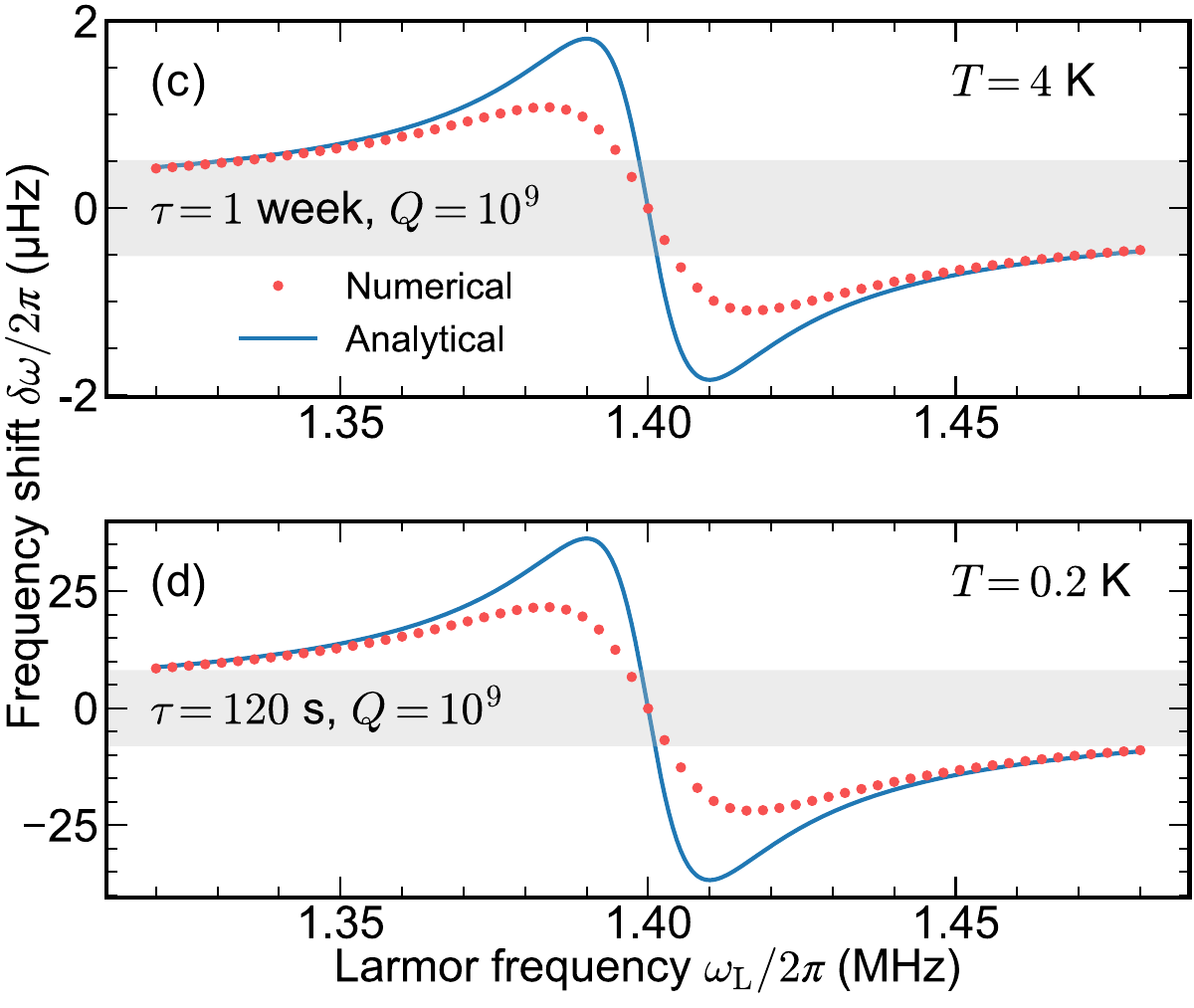}
		\label{fig:membrane_analy_driven}}

	\caption{Exact numerical simulations of a SiN membrane resonator~\cite{Tsaturyan_2017}, $m=\SI{5e-12}{\kilo\gram}$, $\wo/2\pi=\SI{1.4}{\mega\hertz}$. The same resonator is simulated for different temperatures and different driving strengths. For (a) and (b), the drive is chosen at the thermal motion amplitude whereas for (c) and (d) it is chosen to be 2 orders of magnitude above the former. Note the strong decrease in thermal frequency noise in (c) and (d) at the expense of a reduced frequency shift. The drive amplitudes are $z_0=\SI{0.4}{\pico\meter}$ for (a), $z_0=\SI{0.08}{\pico\meter}$ for (b) and $z_0=\SI{10}{\pico\meter}$ for (c) and (d).  Common simulated parameters are $G_x=G_y=\SI{2}{\mega\tesla/\meter}$, $G_z=\SI{1}{\mega\tesla/\meter}$, $N=10^6$ spins, $T_2 = \SI{100}{\micro\second}$, $T_1=\SI{50}{\milli\second}$ and resonant driving of the resonator $\wdd=\wo$.}
	\label{fig:sim_membrane}
\end{figure}

Figure \ref{fig:sim_membrane} shows the analytical (blue lines) and the numerical (red dots) results for a SiN membrane resonator~\cite{Tsaturyan_2017}. Figure \ref{fig:sim_membrane}(a) and (b) show the results for the case with driving amplitude close to the thermal motion of the membrane at the given temperature ($T=\SI{4}{\kelvin}$ for (a) and $T=\SI{0.2}{\kelvin}$ for (b)). In these cases, the numerical data follows very closely the analytical model as the condition $\Omega_R\ll\gp,\frac{1}{T_1}$ is strongly fulfilled. However, the integration time to resolve the frequency shift above the frequency noise is extremely long. On the opposite, Figure \ref{fig:sim_membrane}(c) and (d) show the same case but with increased driving amplitude ($z_0=\SI{10}{\pico\meter}$ for both cases). In this case, $\Omega_R$ is still smaller than $\gp$ but not anymore smaller than $\frac{1}{T_1}$ (in our simulations $T_1=\SI{50}{\milli\second}$). We see that the numerical frequency shift is now smaller than the analytical one, nevertheless, the integration time to resolve the frequency shift is now strongly reduced to feasible values.

\begin{figure}[H]
	\centering
	\includegraphics[width=0.5\textwidth]{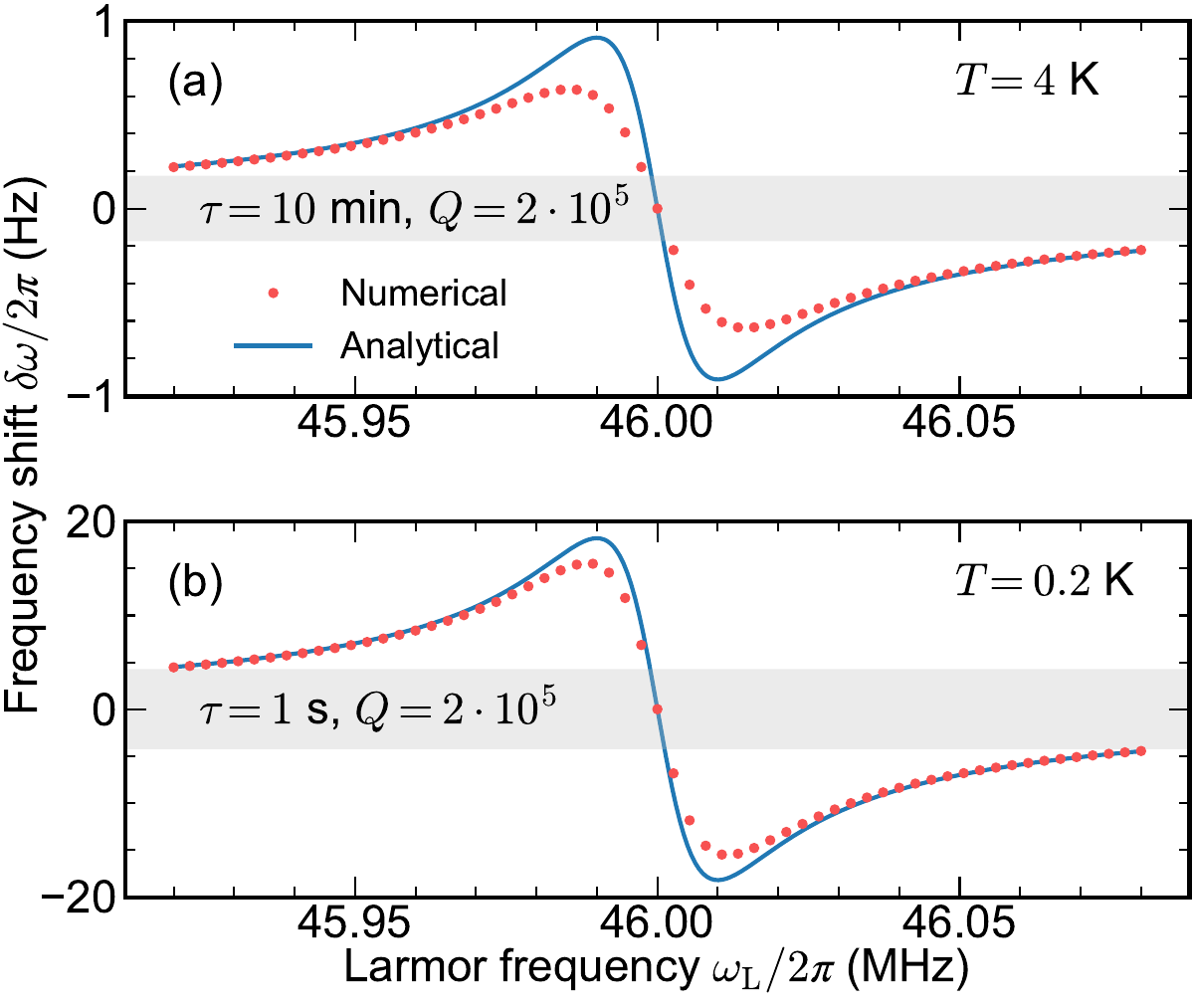}
	\caption{Exact numerical simulations of a graphene sheet resonator~\cite{Weber_2016}, $m=10^{-17}$\,kg, $\wo/2\pi=\SI{46}{\mega\hertz}$, $z_0=\SI{10}{\pico\meter}$ in the (a) $T=\SI{4}{\kelvin}$ case and $z_0=\SI{2}{\pico\meter}$ in the (b) $T=\SI{0.2}{\kelvin}$ case. Common simulated parameters are $G_x=G_y=\SI{2}{\mega\tesla/\meter}$, $G_z=\SI{1}{\mega\tesla/\meter}$, $N=10^6$ spins, $T_2 = \SI{100}{\micro\second}$, $T_1=\SI{50}{\milli\second}$ and resonant driving of the resonator $\wdd=\wo$.}
	\label{fig:graphene_analy}
\end{figure}

From Eq.~\eqref{eq:eom_Ixy_fin_S} we see that the spin dynamics should feature an additional effect that is not described by the analytical model: the jittering of the rotation frequency along the polarization axis ($\Iz$ in our case) due to the modulation of the Larmor frequency from the $G_z$ gradient. The jittering is caused by the second term on the r.h.s. of the equation, which contains a magnetic gradient-dependent contribution. As the oscillator is driven, it is possible to leverage this frequency jittering. For the case of the string resonator presented in the main text, the Larmor frequency is $\wl/2\pi = \SI{5.5}{\mega\hertz}$, the greatest driving strength (at $T=\SI{77}{\kelvin}$) is $z_0=\SI{20}{\pico\meter}$, and the gradient along $z$ is $G_z=\SI{1}{\mega\tesla/\meter}$ MT/m. The maximum amplitude of the frequency modulation (recall that $q$ oscillates) is $\gamma G_zz_0/2\pi\sim\SI{1}{\kilo\hertz}$, amounting to $\sim\SI{0.02}{\percent}$ of the precession frequency, so the effect is very limited for this resonator. For the membrane resonator, the greatest driving is \SI{10}{\pico\meter}, giving a frequency jittering of $\sim\SI{500}{\hertz}$, amounting here to $\sim\SI{0.04}{\percent}$ of the precession frequency. For the third resonator, the graphene sheet, the effect is even smaller due to the higher Larmor frequency: in this case, the jittering is smaller than $\SI{0.001}{\percent}$ of the precession frequency. We conclude that this effect can be ignored for the resonators we are considering.

\subsection{Magnetic tip simulations}\label{sub_sect:magnetic_sim}
To extract a meaningful value for the magnetic field gradients $G_i$, we perform a numerical simulation of the magnetic field of a cobalt nanomagnet. The nanomagnet resembles a cylinder of length $L=\SI{1}{\micro\meter}$ and radius $R=\SI{50}{\nano\meter}$. We are directly inspired by the nanomagnet presented in Ref.~\cite{Longenecker_2012}. We assume that the nanomagnet is pre-magnetized to \SI{1}{\tesla} and we apply an external magnetic field. The latter is used to tune the region where the Larmor frequency matches the mechanical resonator's frequency; we want it to be as close as possible to the nanomagnet in order to harvest the highest magnetic field gradients. Hence, the external magnetic field can be in the opposite direction of the nanomagnet $z$-magnetic field depending on the device investigated, as the required magnetic field for frequency matching can be smaller than the nanomagnet-generated magnetic field. The nanomagnet magnetization should remain roughly constant due to the shape anisotropy, which turns our Co cylinder effectively into a hard magnet~\cite{Longenecker_2012}.

\begin{figure}[H]
	\centering
	\includegraphics[width=0.85\textwidth]{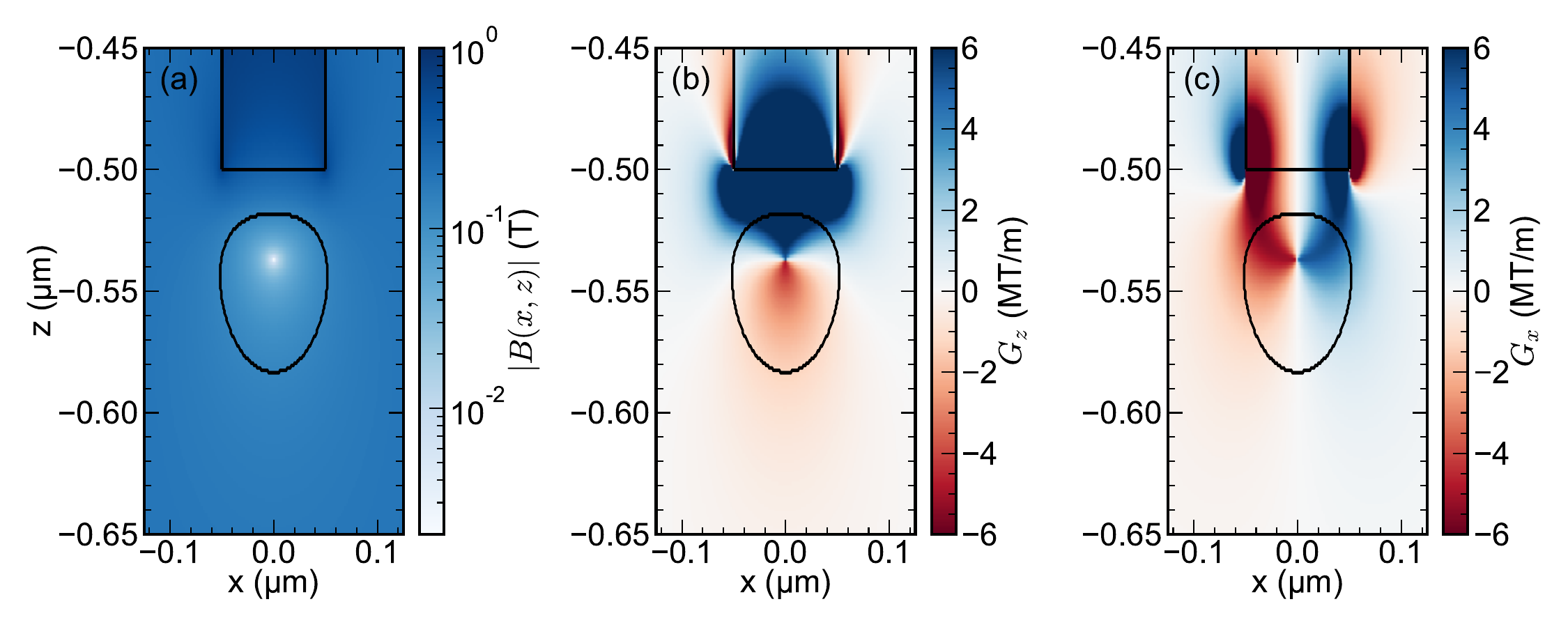}
	\caption{Numerical simulation of a cobalt nanomagnet (black rectangle) pre-magnetized at 1 T subjected to an external magnetic field of \SI{0.2}{\tesla} in the $z$-direction (bottom to top). (a) Absolute value of the magnetic field. The black line shows where the magnetic field is resonant with the mechanical resonator: $\gamma B_0 = \wl = \wo = 2\pi\times\SI{5.5}{\mega\hertz}$. The magnetic field gradients are calculated from (a) and result in a $G_z$ (b) and a $G_x$ (c) component. Note that $G_x$ and $G_z$ are the magnetic field gradients in the $x$ and $z$ directions of the spin reference frame (and not the nanomagnet reference frame).}
	\label{fig:mag_tip_sim_1}
\end{figure}

Figure \ref{fig:mag_tip_sim_1}(a) shows the absolute value of the magnetic field in the vicinity of the nanomagnet (black rectangle) for the case of a SiN string with $\omega_0/2\pi = \SI{5.5}{\mega\hertz}$. In this case, we apply an external magnetic field of \SI{0.2}{\tesla} in the opposite direction of the nanomagnet $z$-magnetic field. The region where the Larmor frequency of the spins would be resonant with the resonator mechanical frequency ($\wl=\wo$) is showed as a black line. We can then extract the magnetic field gradients in the $x$ and $z$ directions of the spin reference frame. These gradients are displayed on Fig.~\ref{fig:mag_tip_sim_1}(b) and \ref{fig:mag_tip_sim_1}(c). In the optimal case, the sample would be in a region where $G_x$ is maximal and $G_z$ minimal. In addition, the sample must be small enough so that it does not overlap the right and left lobes otherwise the effect of the $G_x$ gradient would cancel out due to the sign inversion of the latter. 

From this simulation, we extract the value of the gradients used in the main text, namely $G_x=G_y=\SI{2}{\mega\tesla/\meter}$ ($G_x=G_y$ by symmetry) and $G_z=\SI{1}{\mega\tesla/\meter}$.

\subsection{Boltzmann vs statistical polarization}

\begin{figure}[H]
	\centering
	\includegraphics[width=0.5\textwidth]{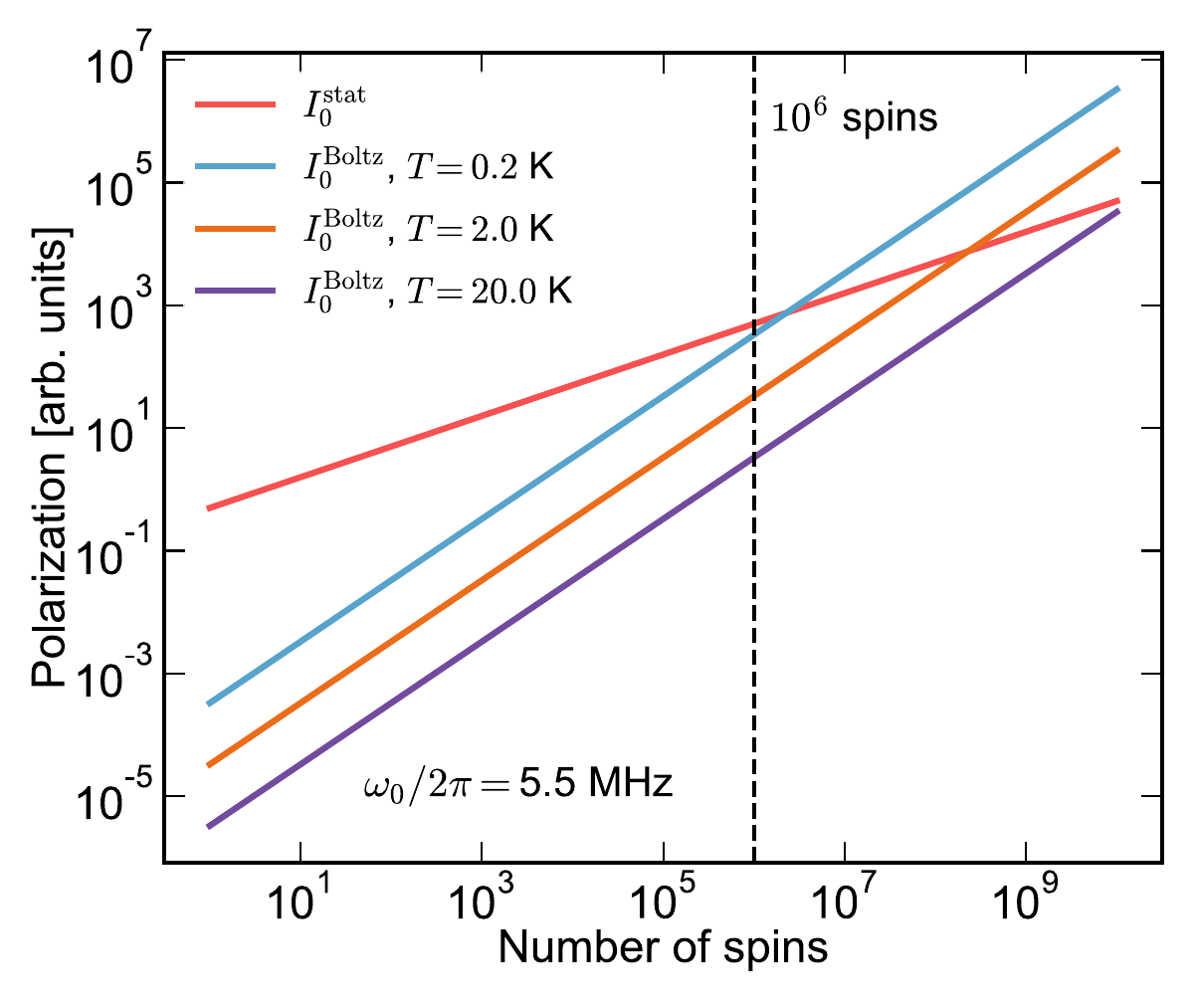}
	\caption{Boltzmann polarization compared to the statistical polarization for different numbers of spins in the sample and different temperatures for the string resonator case. The black dashed line represents a sample of $10^6$ spins like the one mainly used throughout the main text.}
	\label{fig:Boltz_vs_stat_I0}
\end{figure}

To justify the interest in looking at the statistical polarization of the spins instead of the Boltzmann polarization, we can easily plot the different values for a range of temperature and number of spins in the sample. The Boltzmann polarization is given by Eq.~\eqref{eq:Boltzmann_pol} whereas the statistical polarization is given by $I_0 = \frac{1}{2}\sqrt{N}$~\cite{Degen_2007}. The comparison is shown in Fig.~\ref{fig:Boltz_vs_stat_I0} for the string resonator. The black dashed line shows the case of $10^6$ spins used in the main text. It is clear that for samples containing fewer spins the statistical polarization would allow a much stronger signal than the Boltzmann polarization in the same conditions.

\begin{thebibliography}{59}%
	\makeatletter
	\providecommand \@ifxundefined [1]{%
		\@ifx{#1\undefined}
	}%
	\providecommand \@ifnum [1]{%
		\ifnum #1\expandafter \@firstoftwo
		\else \expandafter \@secondoftwo
		\fi
	}%
	\providecommand \@ifx [1]{%
		\ifx #1\expandafter \@firstoftwo
		\else \expandafter \@secondoftwo
		\fi
	}%
	\providecommand \natexlab [1]{#1}%
	\providecommand \enquote  [1]{``#1''}%
	\providecommand \bibnamefont  [1]{#1}%
	\providecommand \bibfnamefont [1]{#1}%
	\providecommand \citenamefont [1]{#1}%
	\providecommand \href@noop [0]{\@secondoftwo}%
	\providecommand \href [0]{\begingroup \@sanitize@url \@href}%
	\providecommand \@href[1]{\@@startlink{#1}\@@href}%
	\providecommand \@@href[1]{\endgroup#1\@@endlink}%
	\providecommand \@sanitize@url [0]{\catcode `\\12\catcode `\$12\catcode
		`\&12\catcode `\#12\catcode `\^12\catcode `\_12\catcode `\%12\relax}%
	\providecommand \@@startlink[1]{}%
	\providecommand \@@endlink[0]{}%
	\providecommand \url  [0]{\begingroup\@sanitize@url \@url }%
	\providecommand \@url [1]{\endgroup\@href {#1}{\urlprefix }}%
	\providecommand \urlprefix  [0]{URL }%
	\providecommand \Eprint [0]{\href }%
	\providecommand \doibase [0]{https://doi.org/}%
	\providecommand \selectlanguage [0]{\@gobble}%
	\providecommand \bibinfo  [0]{\@secondoftwo}%
	\providecommand \bibfield  [0]{\@secondoftwo}%
	\providecommand \translation [1]{[#1]}%
	\providecommand \BibitemOpen [0]{}%
	\providecommand \bibitemStop [0]{}%
	\providecommand \bibitemNoStop [0]{.\EOS\space}%
	\providecommand \EOS [0]{\spacefactor3000\relax}%
	\providecommand \BibitemShut  [1]{\csname bibitem#1\endcsname}%
	\let\auto@bib@innerbib\@empty
	\bibitem [{\citenamefont {Sidles}(1991)}]{Sidles_1991}%
	\BibitemOpen
	\bibfield  {author} {\bibinfo {author} {\bibfnamefont {J.~A.}\ \bibnamefont
			{Sidles}},\ }\bibfield  {title} {\bibinfo {title} {Noninductive detection of
			single-proton magnetic resonance},\ }\href {https://doi.org/10.1063/1.104757}
	{\bibfield  {journal} {\bibinfo  {journal} {Applied Physics Letters}\
		}\textbf {\bibinfo {volume} {58}},\ \bibinfo {pages} {2854} (\bibinfo {year}
		{1991})}\BibitemShut {NoStop}%
	\bibitem [{\citenamefont {Poggio}\ and\ \citenamefont
		{Degen}(2010)}]{Poggio_2010}%
	\BibitemOpen
	\bibfield  {author} {\bibinfo {author} {\bibfnamefont {M.}~\bibnamefont
			{Poggio}}\ and\ \bibinfo {author} {\bibfnamefont {C.~L.}\ \bibnamefont
			{Degen}},\ }\bibfield  {title} {\bibinfo {title} {Force-detected nuclear
			magnetic resonance: recent advances and future challenges},\ }\href
	{https://doi.org/10.1088/0957-4484/21/34/342001} {\bibfield  {journal}
		{\bibinfo  {journal} {Nanotechnology}\ }\textbf {\bibinfo {volume} {21}},\
		\bibinfo {pages} {342001} (\bibinfo {year} {2010})}\BibitemShut {NoStop}%
	\bibitem [{\citenamefont {Degen}\ \emph {et~al.}(2009)\citenamefont {Degen},
		\citenamefont {Poggio}, \citenamefont {Mamin}, \citenamefont {Rettner},\ and\
		\citenamefont {Rugar}}]{Degen_2009}%
	\BibitemOpen
	\bibfield  {author} {\bibinfo {author} {\bibfnamefont {C.~L.}\ \bibnamefont
			{Degen}}, \bibinfo {author} {\bibfnamefont {M.}~\bibnamefont {Poggio}},
		\bibinfo {author} {\bibfnamefont {H.~J.}\ \bibnamefont {Mamin}}, \bibinfo
		{author} {\bibfnamefont {C.~T.}\ \bibnamefont {Rettner}},\ and\ \bibinfo
		{author} {\bibfnamefont {D.}~\bibnamefont {Rugar}},\ }\bibfield  {title}
	{\bibinfo {title} {Nanoscale magnetic resonance imaging},\ }\href
	{https://doi.org/10.1073/pnas.0812068106} {\bibfield  {journal} {\bibinfo
			{journal} {Proceedings of the National Academy of Sciences}\ }\textbf
		{\bibinfo {volume} {106}},\ \bibinfo {pages} {1313} (\bibinfo {year}
		{2009})}\BibitemShut {NoStop}%
	\bibitem [{\citenamefont {Nichol}\ \emph {et~al.}(2013)\citenamefont {Nichol},
		\citenamefont {Naibert}, \citenamefont {Hemesath}, \citenamefont {Lauhon},\
		and\ \citenamefont {Budakian}}]{Nichol_2013}%
	\BibitemOpen
	\bibfield  {author} {\bibinfo {author} {\bibfnamefont {J.~M.}\ \bibnamefont
			{Nichol}}, \bibinfo {author} {\bibfnamefont {T.~R.}\ \bibnamefont {Naibert}},
		\bibinfo {author} {\bibfnamefont {E.~R.}\ \bibnamefont {Hemesath}}, \bibinfo
		{author} {\bibfnamefont {L.~J.}\ \bibnamefont {Lauhon}},\ and\ \bibinfo
		{author} {\bibfnamefont {R.}~\bibnamefont {Budakian}},\ }\bibfield  {title}
	{\bibinfo {title} {Nanoscale fourier-transform magnetic resonance imaging},\
	}\href {https://doi.org/10.1103/PhysRevX.3.031016} {\bibfield  {journal}
		{\bibinfo  {journal} {Physical Review X}\ }\textbf {\bibinfo {volume} {3}},\
		\bibinfo {pages} {031016} (\bibinfo {year} {2013})}\BibitemShut {NoStop}%
	\bibitem [{\citenamefont {Grob}\ \emph {et~al.}(2019)\citenamefont {Grob},
		\citenamefont {Krass}, \citenamefont {Héritier}, \citenamefont {Pachlatko},
		\citenamefont {Rhensius}, \citenamefont {Košata}, \citenamefont {Moores},
		\citenamefont {Takahashi}, \citenamefont {Eichler},\ and\ \citenamefont
		{Degen}}]{Grob_2019}%
	\BibitemOpen
	\bibfield  {author} {\bibinfo {author} {\bibfnamefont {U.}~\bibnamefont
			{Grob}}, \bibinfo {author} {\bibfnamefont {M.~D.}\ \bibnamefont {Krass}},
		\bibinfo {author} {\bibfnamefont {M.}~\bibnamefont {Héritier}}, \bibinfo
		{author} {\bibfnamefont {R.}~\bibnamefont {Pachlatko}}, \bibinfo {author}
		{\bibfnamefont {J.}~\bibnamefont {Rhensius}}, \bibinfo {author}
		{\bibfnamefont {J.}~\bibnamefont {Košata}}, \bibinfo {author} {\bibfnamefont
			{B.~A.}\ \bibnamefont {Moores}}, \bibinfo {author} {\bibfnamefont
			{H.}~\bibnamefont {Takahashi}}, \bibinfo {author} {\bibfnamefont
			{A.}~\bibnamefont {Eichler}},\ and\ \bibinfo {author} {\bibfnamefont {C.~L.}\
			\bibnamefont {Degen}},\ }\bibfield  {title} {\bibinfo {title} {Magnetic
			resonance force microscopy with a one-dimensional resolution of 0.9
			nanometers},\ }\href {https://doi.org/10.1021/acs.nanolett.9b03048}
	{\bibfield  {journal} {\bibinfo  {journal} {Nano Letters}\ }\textbf {\bibinfo
			{volume} {19}},\ \bibinfo {pages} {7935} (\bibinfo {year}
		{2019})}\BibitemShut {NoStop}%
	\bibitem [{\citenamefont {Haas}\ \emph {et~al.}(2022)\citenamefont {Haas},
		\citenamefont {Tabatabaei}, \citenamefont {Rose}, \citenamefont {Sahafi},
		\citenamefont {Piscitelli}, \citenamefont {Jordan}, \citenamefont
		{Priyadarsi}, \citenamefont {Singh}, \citenamefont {Yager}, \citenamefont
		{Poole}, \citenamefont {Dalacu},\ and\ \citenamefont {Budakian}}]{Haas_2022}%
	\BibitemOpen
	\bibfield  {author} {\bibinfo {author} {\bibfnamefont {H.}~\bibnamefont
			{Haas}}, \bibinfo {author} {\bibfnamefont {S.}~\bibnamefont {Tabatabaei}},
		\bibinfo {author} {\bibfnamefont {W.}~\bibnamefont {Rose}}, \bibinfo {author}
		{\bibfnamefont {P.}~\bibnamefont {Sahafi}}, \bibinfo {author} {\bibfnamefont
			{M.}~\bibnamefont {Piscitelli}}, \bibinfo {author} {\bibfnamefont
			{A.}~\bibnamefont {Jordan}}, \bibinfo {author} {\bibfnamefont
			{P.}~\bibnamefont {Priyadarsi}}, \bibinfo {author} {\bibfnamefont
			{N.}~\bibnamefont {Singh}}, \bibinfo {author} {\bibfnamefont
			{B.}~\bibnamefont {Yager}}, \bibinfo {author} {\bibfnamefont {P.~J.}\
			\bibnamefont {Poole}}, \bibinfo {author} {\bibfnamefont {D.}~\bibnamefont
			{Dalacu}},\ and\ \bibinfo {author} {\bibfnamefont {R.}~\bibnamefont
			{Budakian}},\ }\bibfield  {title} {\bibinfo {title} {Nuclear magnetic
			resonance diffraction with subangstrom precision},\ }\href
	{https://doi.org/10.1073/pnas.2209213119} {\bibfield  {journal} {\bibinfo
			{journal} {Proceedings of the National Academy of Sciences}\ }\textbf
		{\bibinfo {volume} {119}},\ \bibinfo {pages} {e2209213119} (\bibinfo {year}
		{2022})}\BibitemShut {NoStop}%
	\bibitem [{\citenamefont {Eichler}(2022)}]{Eichler_2022}%
	\BibitemOpen
	\bibfield  {author} {\bibinfo {author} {\bibfnamefont {A.}~\bibnamefont
			{Eichler}},\ }\bibfield  {title} {\bibinfo {title} {Ultra-high-q
			nanomechanical resonators for force sensing},\ }\href
	{https://doi.org/10.1088/2633-4356/acaba4} {\bibfield  {journal} {\bibinfo
			{journal} {Materials for Quantum Technology}\ }\textbf {\bibinfo {volume}
			{2}},\ \bibinfo {pages} {043001} (\bibinfo {year} {2022})}\BibitemShut
	{NoStop}%
	\bibitem [{\citenamefont {Reinhardt}\ \emph {et~al.}(2016)\citenamefont
		{Reinhardt}, \citenamefont {Müller}, \citenamefont {Bourassa},\ and\
		\citenamefont {Sankey}}]{Reinhardt_2016}%
	\BibitemOpen
	\bibfield  {author} {\bibinfo {author} {\bibfnamefont {C.}~\bibnamefont
			{Reinhardt}}, \bibinfo {author} {\bibfnamefont {T.}~\bibnamefont {Müller}},
		\bibinfo {author} {\bibfnamefont {A.}~\bibnamefont {Bourassa}},\ and\
		\bibinfo {author} {\bibfnamefont {J.~C.}\ \bibnamefont {Sankey}},\ }\bibfield
	{title} {\bibinfo {title} {Ultralow-noise sin trampoline resonators for
			sensing and optomechanics},\ }\href
	{https://doi.org/10.1103/PhysRevX.6.021001} {\bibfield  {journal} {\bibinfo
			{journal} {Physical Review X}\ }\textbf {\bibinfo {volume} {6}},\ \bibinfo
		{pages} {021001} (\bibinfo {year} {2016})}\BibitemShut {NoStop}%
	\bibitem [{\citenamefont {Norte}\ \emph {et~al.}(2016)\citenamefont {Norte},
		\citenamefont {Moura},\ and\ \citenamefont {Gröblacher}}]{Norte_2016}%
	\BibitemOpen
	\bibfield  {author} {\bibinfo {author} {\bibfnamefont {R.}~\bibnamefont
			{Norte}}, \bibinfo {author} {\bibfnamefont {J.}~\bibnamefont {Moura}},\ and\
		\bibinfo {author} {\bibfnamefont {S.}~\bibnamefont {Gröblacher}},\
	}\bibfield  {title} {\bibinfo {title} {Mechanical resonators for quantum
			optomechanics experiments at room temperature},\ }\href
	{https://doi.org/10.1103/PhysRevLett.116.147202} {\bibfield  {journal}
		{\bibinfo  {journal} {Physical Review Letters}\ }\textbf {\bibinfo {volume}
			{116}},\ \bibinfo {pages} {147202} (\bibinfo {year} {2016})}\BibitemShut
	{NoStop}%
	\bibitem [{\citenamefont {Tsaturyan}\ \emph {et~al.}(2017)\citenamefont
		{Tsaturyan}, \citenamefont {Barg}, \citenamefont {Polzik},\ and\
		\citenamefont {Schliesser}}]{Tsaturyan_2017}%
	\BibitemOpen
	\bibfield  {author} {\bibinfo {author} {\bibfnamefont {Y.}~\bibnamefont
			{Tsaturyan}}, \bibinfo {author} {\bibfnamefont {A.}~\bibnamefont {Barg}},
		\bibinfo {author} {\bibfnamefont {E.~S.}\ \bibnamefont {Polzik}},\ and\
		\bibinfo {author} {\bibfnamefont {A.}~\bibnamefont {Schliesser}},\ }\bibfield
	{title} {\bibinfo {title} {Ultracoherent nanomechanical resonators via soft
			clamping and dissipation dilution},\ }\href
	{https://doi.org/10.1038/nnano.2017.101} {\bibfield  {journal} {\bibinfo
			{journal} {Nature Nanotechnology}\ }\textbf {\bibinfo {volume} {12}},\
		\bibinfo {pages} {776} (\bibinfo {year} {2017})}\BibitemShut {NoStop}%
	\bibitem [{\citenamefont {Rossi}\ \emph {et~al.}(2018)\citenamefont {Rossi},
		\citenamefont {Mason}, \citenamefont {Chen}, \citenamefont {Tsaturyan},\ and\
		\citenamefont {Schliesser}}]{Rossi_2018}%
	\BibitemOpen
	\bibfield  {author} {\bibinfo {author} {\bibfnamefont {M.}~\bibnamefont
			{Rossi}}, \bibinfo {author} {\bibfnamefont {D.}~\bibnamefont {Mason}},
		\bibinfo {author} {\bibfnamefont {J.}~\bibnamefont {Chen}}, \bibinfo {author}
		{\bibfnamefont {Y.}~\bibnamefont {Tsaturyan}},\ and\ \bibinfo {author}
		{\bibfnamefont {A.}~\bibnamefont {Schliesser}},\ }\bibfield  {title}
	{\bibinfo {title} {Measurement-based quantum control of mechanical motion},\
	}\href {https://doi.org/10.1038/s41586-018-0643-8} {\bibfield  {journal}
		{\bibinfo  {journal} {Nature}\ }\textbf {\bibinfo {volume} {563}},\ \bibinfo
		{pages} {53} (\bibinfo {year} {2018})}\BibitemShut {NoStop}%
	\bibitem [{\citenamefont {Reetz}\ \emph {et~al.}(2019)\citenamefont {Reetz},
		\citenamefont {Fischer}, \citenamefont {Assumpção}, \citenamefont
		{McNally}, \citenamefont {Burns}, \citenamefont {Sankey},\ and\ \citenamefont
		{Regal}}]{Reetz_2019}%
	\BibitemOpen
	\bibfield  {author} {\bibinfo {author} {\bibfnamefont {C.}~\bibnamefont
			{Reetz}}, \bibinfo {author} {\bibfnamefont {R.}~\bibnamefont {Fischer}},
		\bibinfo {author} {\bibfnamefont {G.}~\bibnamefont {Assumpção}}, \bibinfo
		{author} {\bibfnamefont {D.}~\bibnamefont {McNally}}, \bibinfo {author}
		{\bibfnamefont {P.}~\bibnamefont {Burns}}, \bibinfo {author} {\bibfnamefont
			{J.}~\bibnamefont {Sankey}},\ and\ \bibinfo {author} {\bibfnamefont
			{C.}~\bibnamefont {Regal}},\ }\bibfield  {title} {\bibinfo {title} {Analysis
			of membrane phononic crystals with wide band gaps and low-mass defects},\
	}\href {https://doi.org/10.1103/physrevapplied.12.044027} {\bibfield
		{journal} {\bibinfo  {journal} {Physical Review Applied}\ }\textbf {\bibinfo
			{volume} {12}},\ \bibinfo {pages} {044027} (\bibinfo {year}
		{2019})}\BibitemShut {NoStop}%
	\bibitem [{\citenamefont {Ghadimi}\ \emph {et~al.}(2018)\citenamefont
		{Ghadimi}, \citenamefont {Fedorov}, \citenamefont {Engelsen}, \citenamefont
		{Bereyhi}, \citenamefont {Schilling}, \citenamefont {Wilson},\ and\
		\citenamefont {Kippenberg}}]{Ghadimi_2018}%
	\BibitemOpen
	\bibfield  {author} {\bibinfo {author} {\bibfnamefont {A.~H.}\ \bibnamefont
			{Ghadimi}}, \bibinfo {author} {\bibfnamefont {S.~A.}\ \bibnamefont
			{Fedorov}}, \bibinfo {author} {\bibfnamefont {N.~J.}\ \bibnamefont
			{Engelsen}}, \bibinfo {author} {\bibfnamefont {M.~J.}\ \bibnamefont
			{Bereyhi}}, \bibinfo {author} {\bibfnamefont {R.}~\bibnamefont {Schilling}},
		\bibinfo {author} {\bibfnamefont {D.~J.}\ \bibnamefont {Wilson}},\ and\
		\bibinfo {author} {\bibfnamefont {T.~J.}\ \bibnamefont {Kippenberg}},\
	}\bibfield  {title} {\bibinfo {title} {Elastic strain engineering for
			ultralow mechanical dissipation},\ }\href
	{https://doi.org/10.1126/science.aar6939} {\bibfield  {journal} {\bibinfo
			{journal} {Science}\ }\textbf {\bibinfo {volume} {360}},\ \bibinfo {pages}
		{764} (\bibinfo {year} {2018})}\BibitemShut {NoStop}%
	\bibitem [{\citenamefont {Beccari}\ \emph {et~al.}(2022)\citenamefont
		{Beccari}, \citenamefont {Visani}, \citenamefont {Fedorov}, \citenamefont
		{Bereyhi}, \citenamefont {Boureau}, \citenamefont {Engelsen},\ and\
		\citenamefont {Kippenberg}}]{Beccari_2022}%
	\BibitemOpen
	\bibfield  {author} {\bibinfo {author} {\bibfnamefont {A.}~\bibnamefont
			{Beccari}}, \bibinfo {author} {\bibfnamefont {D.~A.}\ \bibnamefont {Visani}},
		\bibinfo {author} {\bibfnamefont {S.~A.}\ \bibnamefont {Fedorov}}, \bibinfo
		{author} {\bibfnamefont {M.~J.}\ \bibnamefont {Bereyhi}}, \bibinfo {author}
		{\bibfnamefont {V.}~\bibnamefont {Boureau}}, \bibinfo {author} {\bibfnamefont
			{N.~J.}\ \bibnamefont {Engelsen}},\ and\ \bibinfo {author} {\bibfnamefont
			{T.~J.}\ \bibnamefont {Kippenberg}},\ }\bibfield  {title} {\bibinfo {title}
		{Strained crystalline nanomechanical resonators with quality factors above 10
			billion},\ }\href {https://doi.org/10.1038/s41567-021-01498-4} {\bibfield
		{journal} {\bibinfo  {journal} {Nature Physics}\ }\textbf {\bibinfo {volume}
			{18}},\ \bibinfo {pages} {436} (\bibinfo {year} {2022})}\BibitemShut
	{NoStop}%
	\bibitem [{\citenamefont {Gisler}\ \emph {et~al.}(2022)\citenamefont {Gisler},
		\citenamefont {Helal}, \citenamefont {Sabonis}, \citenamefont {Grob},
		\citenamefont {Héritier}, \citenamefont {Degen}, \citenamefont {Ghadimi},\
		and\ \citenamefont {Eichler}}]{Gisler_2022}%
	\BibitemOpen
	\bibfield  {author} {\bibinfo {author} {\bibfnamefont {T.}~\bibnamefont
			{Gisler}}, \bibinfo {author} {\bibfnamefont {M.}~\bibnamefont {Helal}},
		\bibinfo {author} {\bibfnamefont {D.}~\bibnamefont {Sabonis}}, \bibinfo
		{author} {\bibfnamefont {U.}~\bibnamefont {Grob}}, \bibinfo {author}
		{\bibfnamefont {M.}~\bibnamefont {Héritier}}, \bibinfo {author}
		{\bibfnamefont {C.~L.}\ \bibnamefont {Degen}}, \bibinfo {author}
		{\bibfnamefont {A.~H.}\ \bibnamefont {Ghadimi}},\ and\ \bibinfo {author}
		{\bibfnamefont {A.}~\bibnamefont {Eichler}},\ }\bibfield  {title} {\bibinfo
		{title} {Soft-clamped silicon nitride string resonators at millikelvin
			temperatures},\ }\href {https://doi.org/10.1103/PhysRevLett.129.104301}
	{\bibfield  {journal} {\bibinfo  {journal} {Physical Review Letters}\
		}\textbf {\bibinfo {volume} {129}},\ \bibinfo {pages} {104301} (\bibinfo
		{year} {2022})}\BibitemShut {NoStop}%
	\bibitem [{\citenamefont {Bereyhi}\ \emph
		{et~al.}(2022{\natexlab{a}})\citenamefont {Bereyhi}, \citenamefont
		{Arabmoheghi}, \citenamefont {Beccari}, \citenamefont {Fedorov},
		\citenamefont {Huang}, \citenamefont {Kippenberg},\ and\ \citenamefont
		{Engelsen}}]{Bereyhi_2022}%
	\BibitemOpen
	\bibfield  {author} {\bibinfo {author} {\bibfnamefont {M.~J.}\ \bibnamefont
			{Bereyhi}}, \bibinfo {author} {\bibfnamefont {A.}~\bibnamefont
			{Arabmoheghi}}, \bibinfo {author} {\bibfnamefont {A.}~\bibnamefont
			{Beccari}}, \bibinfo {author} {\bibfnamefont {S.~A.}\ \bibnamefont
			{Fedorov}}, \bibinfo {author} {\bibfnamefont {G.}~\bibnamefont {Huang}},
		\bibinfo {author} {\bibfnamefont {T.~J.}\ \bibnamefont {Kippenberg}},\ and\
		\bibinfo {author} {\bibfnamefont {N.~J.}\ \bibnamefont {Engelsen}},\
	}\bibfield  {title} {\bibinfo {title} {Perimeter modes of nanomechanical
			resonators exhibit quality factors exceeding $10^9$ at room temperature},\
	}\href {https://doi.org/10.1103/PhysRevX.12.021036} {\bibfield  {journal}
		{\bibinfo  {journal} {Physical Review X}\ }\textbf {\bibinfo {volume} {12}},\
		\bibinfo {pages} {021036} (\bibinfo {year} {2022}{\natexlab{a}})}\BibitemShut
	{NoStop}%
	\bibitem [{\citenamefont {Fedorov}\ \emph {et~al.}(2020)\citenamefont
		{Fedorov}, \citenamefont {Beccari}, \citenamefont {Engelsen},\ and\
		\citenamefont {Kippenberg}}]{Fedorov_2020}%
	\BibitemOpen
	\bibfield  {author} {\bibinfo {author} {\bibfnamefont {S.}~\bibnamefont
			{Fedorov}}, \bibinfo {author} {\bibfnamefont {A.}~\bibnamefont {Beccari}},
		\bibinfo {author} {\bibfnamefont {N.}~\bibnamefont {Engelsen}},\ and\
		\bibinfo {author} {\bibfnamefont {T.}~\bibnamefont {Kippenberg}},\ }\bibfield
	{title} {\bibinfo {title} {Fractal-like mechanical resonators with a
			soft-clamped fundamental mode},\ }\href
	{https://doi.org/10.1103/PhysRevLett.124.025502} {\bibfield  {journal}
		{\bibinfo  {journal} {Physical Review Letters}\ }\textbf {\bibinfo {volume}
			{124}},\ \bibinfo {pages} {025502} (\bibinfo {year} {2020})}\BibitemShut
	{NoStop}%
	\bibitem [{\citenamefont {Bereyhi}\ \emph
		{et~al.}(2022{\natexlab{b}})\citenamefont {Bereyhi}, \citenamefont {Beccari},
		\citenamefont {Groth}, \citenamefont {Fedorov}, \citenamefont {Arabmoheghi},
		\citenamefont {Kippenberg},\ and\ \citenamefont
		{Engelsen}}]{Bereyhi_2022_NC}%
	\BibitemOpen
	\bibfield  {author} {\bibinfo {author} {\bibfnamefont {M.~J.}\ \bibnamefont
			{Bereyhi}}, \bibinfo {author} {\bibfnamefont {A.}~\bibnamefont {Beccari}},
		\bibinfo {author} {\bibfnamefont {R.}~\bibnamefont {Groth}}, \bibinfo
		{author} {\bibfnamefont {S.~A.}\ \bibnamefont {Fedorov}}, \bibinfo {author}
		{\bibfnamefont {A.}~\bibnamefont {Arabmoheghi}}, \bibinfo {author}
		{\bibfnamefont {T.~J.}\ \bibnamefont {Kippenberg}},\ and\ \bibinfo {author}
		{\bibfnamefont {N.~J.}\ \bibnamefont {Engelsen}},\ }\bibfield  {title}
	{\bibinfo {title} {Hierarchical tensile structures with ultralow mechanical
			dissipation},\ }\href {https://doi.org/10.1038/s41467-022-30586-z} {\bibfield
		{journal} {\bibinfo  {journal} {Nature Communications}\ }\textbf {\bibinfo
			{volume} {13}},\ \bibinfo {pages} {3097} (\bibinfo {year}
		{2022}{\natexlab{b}})}\BibitemShut {NoStop}%
	\bibitem [{\citenamefont {Shin}\ \emph {et~al.}(2021)\citenamefont {Shin},
		\citenamefont {Cupertino}, \citenamefont {de~Jong}, \citenamefont
		{Steeneken}, \citenamefont {Bessa},\ and\ \citenamefont {Norte}}]{Shin_2021}%
	\BibitemOpen
	\bibfield  {author} {\bibinfo {author} {\bibfnamefont {D.}~\bibnamefont
			{Shin}}, \bibinfo {author} {\bibfnamefont {A.}~\bibnamefont {Cupertino}},
		\bibinfo {author} {\bibfnamefont {M.~H.~J.}\ \bibnamefont {de~Jong}},
		\bibinfo {author} {\bibfnamefont {P.~G.}\ \bibnamefont {Steeneken}}, \bibinfo
		{author} {\bibfnamefont {M.~A.}\ \bibnamefont {Bessa}},\ and\ \bibinfo
		{author} {\bibfnamefont {R.~A.}\ \bibnamefont {Norte}},\ }\bibfield  {title}
	{\bibinfo {title} {Spiderweb nanomechanical resonators via bayesian
			optimization: Inspired by nature and guided by machine learning},\ }\href
	{https://doi.org/10.1002/adma.202106248} {\bibfield  {journal} {\bibinfo
			{journal} {Advanced Materials}\ }\textbf {\bibinfo {volume} {34}},\ \bibinfo
		{pages} {2270019} (\bibinfo {year} {2021})}\BibitemShut {NoStop}%
	\bibitem [{\citenamefont {Moser}\ \emph {et~al.}(2013)\citenamefont {Moser},
		\citenamefont {Güttinger}, \citenamefont {Eichler}, \citenamefont
		{Esplandiu}, \citenamefont {Liu}, \citenamefont {Dykman},\ and\ \citenamefont
		{Bachtold}}]{Moser_2013}%
	\BibitemOpen
	\bibfield  {author} {\bibinfo {author} {\bibfnamefont {J.}~\bibnamefont
			{Moser}}, \bibinfo {author} {\bibfnamefont {J.}~\bibnamefont {Güttinger}},
		\bibinfo {author} {\bibfnamefont {A.}~\bibnamefont {Eichler}}, \bibinfo
		{author} {\bibfnamefont {M.~J.}\ \bibnamefont {Esplandiu}}, \bibinfo {author}
		{\bibfnamefont {D.~E.}\ \bibnamefont {Liu}}, \bibinfo {author} {\bibfnamefont
			{M.~I.}\ \bibnamefont {Dykman}},\ and\ \bibinfo {author} {\bibfnamefont
			{A.}~\bibnamefont {Bachtold}},\ }\bibfield  {title} {\bibinfo {title}
		{Ultrasensitive force detection with a nanotube mechanical resonator},\
	}\href {https://doi.org/10.1038/NNANO.2013.97} {\bibfield  {journal}
		{\bibinfo  {journal} {Nature Nanotechnology}\ }\textbf {\bibinfo {volume}
			{8}},\ \bibinfo {pages} {493} (\bibinfo {year} {2013})}\BibitemShut {NoStop}%
	\bibitem [{\citenamefont {Rossi}\ \emph {et~al.}(2016)\citenamefont {Rossi},
		\citenamefont {Braakman}, \citenamefont {Cadeddu}, \citenamefont {Vasyukov},
		\citenamefont {Tütüncüoglu}, \citenamefont {Fontcuberta~i Morral},\ and\
		\citenamefont {Poggio}}]{Rossi_2016}%
	\BibitemOpen
	\bibfield  {author} {\bibinfo {author} {\bibfnamefont {N.}~\bibnamefont
			{Rossi}}, \bibinfo {author} {\bibfnamefont {F.~R.}\ \bibnamefont {Braakman}},
		\bibinfo {author} {\bibfnamefont {D.}~\bibnamefont {Cadeddu}}, \bibinfo
		{author} {\bibfnamefont {D.}~\bibnamefont {Vasyukov}}, \bibinfo {author}
		{\bibfnamefont {G.}~\bibnamefont {Tütüncüoglu}}, \bibinfo {author}
		{\bibfnamefont {A.}~\bibnamefont {Fontcuberta~i Morral}},\ and\ \bibinfo
		{author} {\bibfnamefont {M.}~\bibnamefont {Poggio}},\ }\bibfield  {title}
	{\bibinfo {title} {Vectorial scanning force microscopy using a nanowire
			sensor},\ }\href {https://doi.org/10.1038/nnano.2016.189} {\bibfield
		{journal} {\bibinfo  {journal} {Nature Nanotechnology}\ }\textbf {\bibinfo
			{volume} {12}},\ \bibinfo {pages} {150} (\bibinfo {year} {2016})}\BibitemShut
	{NoStop}%
	\bibitem [{\citenamefont {Hälg}\ \emph {et~al.}(2021)\citenamefont {Hälg},
		\citenamefont {Gisler}, \citenamefont {Tsaturyan}, \citenamefont {Catalini},
		\citenamefont {Grob}, \citenamefont {Krass}, \citenamefont {Héritier},
		\citenamefont {Mattiat}, \citenamefont {Thamm}, \citenamefont {Schirhagl},
		\citenamefont {Langman}, \citenamefont {Schliesser}, \citenamefont {Degen},\
		and\ \citenamefont {Eichler}}]{Halg_2021}%
	\BibitemOpen
	\bibfield  {author} {\bibinfo {author} {\bibfnamefont {D.}~\bibnamefont
			{Hälg}}, \bibinfo {author} {\bibfnamefont {T.}~\bibnamefont {Gisler}},
		\bibinfo {author} {\bibfnamefont {Y.}~\bibnamefont {Tsaturyan}}, \bibinfo
		{author} {\bibfnamefont {L.}~\bibnamefont {Catalini}}, \bibinfo {author}
		{\bibfnamefont {U.}~\bibnamefont {Grob}}, \bibinfo {author} {\bibfnamefont
			{M.-D.}\ \bibnamefont {Krass}}, \bibinfo {author} {\bibfnamefont
			{M.}~\bibnamefont {Héritier}}, \bibinfo {author} {\bibfnamefont
			{H.}~\bibnamefont {Mattiat}}, \bibinfo {author} {\bibfnamefont {A.-K.}\
			\bibnamefont {Thamm}}, \bibinfo {author} {\bibfnamefont {R.}~\bibnamefont
			{Schirhagl}}, \bibinfo {author} {\bibfnamefont {E.~C.}\ \bibnamefont
			{Langman}}, \bibinfo {author} {\bibfnamefont {A.}~\bibnamefont {Schliesser}},
		\bibinfo {author} {\bibfnamefont {C.~L.}\ \bibnamefont {Degen}},\ and\
		\bibinfo {author} {\bibfnamefont {A.}~\bibnamefont {Eichler}},\ }\bibfield
	{title} {\bibinfo {title} {Membrane-based scanning force microscopy},\ }\href
	{https://doi.org/10.1103/PhysRevApplied.15.L021001} {\bibfield  {journal}
		{\bibinfo  {journal} {Physical Review Applied}\ }\textbf {\bibinfo {volume}
			{15}},\ \bibinfo {pages} {l021001} (\bibinfo {year} {2021})}\BibitemShut
	{NoStop}%
	\bibitem [{\citenamefont {Košata}\ \emph {et~al.}(2020)\citenamefont
		{Košata}, \citenamefont {Zilberberg}, \citenamefont {Degen}, \citenamefont
		{Chitra},\ and\ \citenamefont {Eichler}}]{Kosata_2020}%
	\BibitemOpen
	\bibfield  {author} {\bibinfo {author} {\bibfnamefont {J.}~\bibnamefont
			{Košata}}, \bibinfo {author} {\bibfnamefont {O.}~\bibnamefont {Zilberberg}},
		\bibinfo {author} {\bibfnamefont {C.~L.}\ \bibnamefont {Degen}}, \bibinfo
		{author} {\bibfnamefont {R.}~\bibnamefont {Chitra}},\ and\ \bibinfo {author}
		{\bibfnamefont {A.}~\bibnamefont {Eichler}},\ }\bibfield  {title} {\bibinfo
		{title} {Spin detection via parametric frequency conversion in a membrane
			resonator},\ }\href {https://doi.org/10.1103/PhysRevApplied.14.014042}
	{\bibfield  {journal} {\bibinfo  {journal} {Physical Review Applied}\
		}\textbf {\bibinfo {volume} {14}},\ \bibinfo {pages} {014042} (\bibinfo
		{year} {2020})}\BibitemShut {NoStop}%
	\bibitem [{\citenamefont {Aspelmeyer}\ \emph {et~al.}(2014)\citenamefont
		{Aspelmeyer}, \citenamefont {Kippenberg},\ and\ \citenamefont
		{Marquardt}}]{Aspelmeyer_2014}%
	\BibitemOpen
	\bibfield  {author} {\bibinfo {author} {\bibfnamefont {M.}~\bibnamefont
			{Aspelmeyer}}, \bibinfo {author} {\bibfnamefont {T.~J.}\ \bibnamefont
			{Kippenberg}},\ and\ \bibinfo {author} {\bibfnamefont {F.}~\bibnamefont
			{Marquardt}},\ }\bibfield  {title} {\bibinfo {title} {Cavity optomechanics},\
	}\href {https://doi.org/10.1103/RevModPhys.86.1391} {\bibfield  {journal}
		{\bibinfo  {journal} {Reviews of Modern Physics}\ }\textbf {\bibinfo {volume}
			{86}},\ \bibinfo {pages} {1391} (\bibinfo {year} {2014})}\BibitemShut
	{NoStop}%
	\bibitem [{\citenamefont {Bachtold}\ \emph {et~al.}(2022)\citenamefont
		{Bachtold}, \citenamefont {Moser},\ and\ \citenamefont
		{Dykman}}]{RevModPhys.94.045005}%
	\BibitemOpen
	\bibfield  {author} {\bibinfo {author} {\bibfnamefont {A.}~\bibnamefont
			{Bachtold}}, \bibinfo {author} {\bibfnamefont {J.}~\bibnamefont {Moser}},\
		and\ \bibinfo {author} {\bibfnamefont {M.~I.}\ \bibnamefont {Dykman}},\
	}\bibfield  {title} {\bibinfo {title} {Mesoscopic physics of nanomechanical
			systems},\ }\href {https://doi.org/10.1103/RevModPhys.94.045005} {\bibfield
		{journal} {\bibinfo  {journal} {Rev. Mod. Phys.}\ }\textbf {\bibinfo {volume}
			{94}},\ \bibinfo {pages} {045005} (\bibinfo {year} {2022})}\BibitemShut
	{NoStop}%
	\bibitem [{\citenamefont {Sidles}(1992)}]{Sidles_1992}%
	\BibitemOpen
	\bibfield  {author} {\bibinfo {author} {\bibfnamefont {J.~A.}\ \bibnamefont
			{Sidles}},\ }\bibfield  {title} {\bibinfo {title} {Folded stern-gerlach
			experiment as a means for detecting nuclear magnetic resonance in individual
			nuclei},\ }\href {https://doi.org/10.1103/PhysRevLett.68.1124} {\bibfield
		{journal} {\bibinfo  {journal} {Physical Review Letters}\ }\textbf {\bibinfo
			{volume} {68}},\ \bibinfo {pages} {1124} (\bibinfo {year}
		{1992})}\BibitemShut {NoStop}%
	\bibitem [{\citenamefont {Berman}\ and\ \citenamefont
		{Tsifrinovich}(2022)}]{Berman_2022}%
	\BibitemOpen
	\bibfield  {author} {\bibinfo {author} {\bibfnamefont {G.~P.}\ \bibnamefont
			{Berman}}\ and\ \bibinfo {author} {\bibfnamefont {V.~I.}\ \bibnamefont
			{Tsifrinovich}},\ }\bibfield  {title} {\bibinfo {title} {Magnetic resonance
			force microscopy with matching frequencies of cantilever and spin},\ }\href
	{https://doi.org/10.1063/5.0073237} {\bibfield  {journal} {\bibinfo
			{journal} {Journal of Applied Physics}\ }\textbf {\bibinfo {volume} {131}},\
		\bibinfo {pages} {044301} (\bibinfo {year} {2022})}\BibitemShut {NoStop}%
	\bibitem [{\citenamefont {Lee}\ \emph {et~al.}(2016)\citenamefont {Lee},
		\citenamefont {Lee}, \citenamefont {Ovartchaiyapong}, \citenamefont
		{Minguzzi}, \citenamefont {Maze},\ and\ \citenamefont
		{Bleszynski~Jayich}}]{Lee_2016}%
	\BibitemOpen
	\bibfield  {author} {\bibinfo {author} {\bibfnamefont {K.~W.}\ \bibnamefont
			{Lee}}, \bibinfo {author} {\bibfnamefont {D.}~\bibnamefont {Lee}}, \bibinfo
		{author} {\bibfnamefont {P.}~\bibnamefont {Ovartchaiyapong}}, \bibinfo
		{author} {\bibfnamefont {J.}~\bibnamefont {Minguzzi}}, \bibinfo {author}
		{\bibfnamefont {J.~R.}\ \bibnamefont {Maze}},\ and\ \bibinfo {author}
		{\bibfnamefont {A.~C.}\ \bibnamefont {Bleszynski~Jayich}},\ }\bibfield
	{title} {\bibinfo {title} {Strain coupling of a mechanical resonator to a
			single quantum emitter in diamond},\ }\href
	{https://doi.org/10.1103/PhysRevApplied.6.034005} {\bibfield  {journal}
		{\bibinfo  {journal} {Physical Review Applied}\ }\textbf {\bibinfo {volume}
			{6}},\ \bibinfo {pages} {034005} (\bibinfo {year} {2016})}\BibitemShut
	{NoStop}%
	\bibitem [{\citenamefont {Teissier}\ \emph {et~al.}(2017)\citenamefont
		{Teissier}, \citenamefont {Barfuss},\ and\ \citenamefont
		{Maletinsky}}]{Teissier_2017}%
	\BibitemOpen
	\bibfield  {author} {\bibinfo {author} {\bibfnamefont {J.}~\bibnamefont
			{Teissier}}, \bibinfo {author} {\bibfnamefont {A.}~\bibnamefont {Barfuss}},\
		and\ \bibinfo {author} {\bibfnamefont {P.}~\bibnamefont {Maletinsky}},\
	}\bibfield  {title} {\bibinfo {title} {Hybrid continuous dynamical
			decoupling: a photon-phonon doubly dressed spin},\ }\href
	{https://doi.org/10.1088/2040-8986/aa5f62} {\bibfield  {journal} {\bibinfo
			{journal} {Journal of Optics}\ }\textbf {\bibinfo {volume} {19}},\ \bibinfo
		{pages} {044003} (\bibinfo {year} {2017})}\BibitemShut {NoStop}%
	\bibitem [{\citenamefont {MacQuarrie}\ \emph {et~al.}(2013)\citenamefont
		{MacQuarrie}, \citenamefont {Gosavi}, \citenamefont {Jungwirth},
		\citenamefont {Bhave},\ and\ \citenamefont {Fuchs}}]{MacQuarrie_2013}%
	\BibitemOpen
	\bibfield  {author} {\bibinfo {author} {\bibfnamefont {E.~R.}\ \bibnamefont
			{MacQuarrie}}, \bibinfo {author} {\bibfnamefont {T.~A.}\ \bibnamefont
			{Gosavi}}, \bibinfo {author} {\bibfnamefont {N.~R.}\ \bibnamefont
			{Jungwirth}}, \bibinfo {author} {\bibfnamefont {S.~A.}\ \bibnamefont
			{Bhave}},\ and\ \bibinfo {author} {\bibfnamefont {G.~D.}\ \bibnamefont
			{Fuchs}},\ }\bibfield  {title} {\bibinfo {title} {Mechanical spin control of
			nitrogen-vacancy centers in diamond},\ }\href
	{https://doi.org/10.1103/PhysRevLett.111.227602} {\bibfield  {journal}
		{\bibinfo  {journal} {Physical Review Letters}\ }\textbf {\bibinfo {volume}
			{111}},\ \bibinfo {pages} {227602} (\bibinfo {year} {2013})}\BibitemShut
	{NoStop}%
	\bibitem [{Sup()}]{Supplement}%
	\BibitemOpen
	\href@noop {} {\bibinfo {title} {See supplemental material at [url will be
			inserted by publisher] for further details on the analytical approach and
			additional simulation case studies.}}\BibitemShut {Stop}%
	\bibitem [{\citenamefont {Niinikoski}(2020)}]{niinikoski_2020}%
	\BibitemOpen
	\bibfield  {author} {\bibinfo {author} {\bibfnamefont {T.~O.}\ \bibnamefont
			{Niinikoski}},\ }\href {https://doi.org/10.1017/9781108567435} {\emph
		{\bibinfo {title} {The Physics of Polarized Targets}}}\ (\bibinfo
	{publisher} {Cambridge University Press},\ \bibinfo {year}
	{2020})\BibitemShut {NoStop}%
	\bibitem [{\citenamefont {Bloch}(1946)}]{Bloch_1946}%
	\BibitemOpen
	\bibfield  {author} {\bibinfo {author} {\bibfnamefont {F.}~\bibnamefont
			{Bloch}},\ }\bibfield  {title} {\bibinfo {title} {Nuclear induction},\ }\href
	{https://doi.org/10.1103/PhysRev.70.460} {\bibfield  {journal} {\bibinfo
			{journal} {Physical Review}\ }\textbf {\bibinfo {volume} {70}},\ \bibinfo
		{pages} {460} (\bibinfo {year} {1946})}\BibitemShut {NoStop}%
	\bibitem [{\citenamefont {Krack}\ and\ \citenamefont
		{Gross}(2019)}]{Krack_2019}%
	\BibitemOpen
	\bibfield  {author} {\bibinfo {author} {\bibfnamefont {M.}~\bibnamefont
			{Krack}}\ and\ \bibinfo {author} {\bibfnamefont {J.}~\bibnamefont {Gross}},\
	}\href {https://doi.org/10.1007/978-3-030-14023-6} {\emph {\bibinfo {title}
			{Harmonic Balance for Nonlinear Vibration Problems}}}\ (\bibinfo  {publisher}
	{Springer International Publishing},\ \bibinfo {year} {2019})\BibitemShut
	{NoStop}%
	\bibitem [{\citenamefont {Košata}\ \emph {et~al.}(2022)\citenamefont
		{Košata}, \citenamefont {del Pino}, \citenamefont {Heugel},\ and\
		\citenamefont {Zilberberg}}]{Kosata_2022_SP}%
	\BibitemOpen
	\bibfield  {author} {\bibinfo {author} {\bibfnamefont {J.}~\bibnamefont
			{Košata}}, \bibinfo {author} {\bibfnamefont {J.}~\bibnamefont {del Pino}},
		\bibinfo {author} {\bibfnamefont {T.~L.}\ \bibnamefont {Heugel}},\ and\
		\bibinfo {author} {\bibfnamefont {O.}~\bibnamefont {Zilberberg}},\ }\bibfield
	{title} {\bibinfo {title} {Harmonicbalance.jl: A julia suite for nonlinear
			dynamics using harmonic balance},\ }\href
	{https://doi.org/10.21468/SciPostPhysCodeb.6} {\bibfield  {journal} {\bibinfo
			{journal} {SciPost Physics Codebases}\ ,\ \bibinfo {pages} {6}} (\bibinfo
		{year} {2022})}\BibitemShut {NoStop}%
	\bibitem [{\citenamefont {Hairer}\ \emph {et~al.}(1993)\citenamefont {Hairer},
		\citenamefont {Wanner},\ and\ \citenamefont {Nørsett}}]{Hairer_1993}%
	\BibitemOpen
	\bibfield  {author} {\bibinfo {author} {\bibfnamefont {E.}~\bibnamefont
			{Hairer}}, \bibinfo {author} {\bibfnamefont {G.}~\bibnamefont {Wanner}},\
		and\ \bibinfo {author} {\bibfnamefont {S.~P.}\ \bibnamefont {Nørsett}},\
	}\href {https://doi.org/10.1007/978-3-540-78862-1} {\emph {\bibinfo {title}
			{Solving Ordinary Differential Equations I}}}\ (\bibinfo  {publisher}
	{Springer Berlin Heidelberg},\ \bibinfo {year} {1993})\BibitemShut {NoStop}%
	\bibitem [{\citenamefont {Allan}\ \emph {et~al.}(1972)\citenamefont {Allan},
		\citenamefont {Gray},\ and\ \citenamefont {Machlan}}]{Allan_1972}%
	\BibitemOpen
	\bibfield  {author} {\bibinfo {author} {\bibfnamefont {D.~W.}\ \bibnamefont
			{Allan}}, \bibinfo {author} {\bibfnamefont {J.~E.}\ \bibnamefont {Gray}},\
		and\ \bibinfo {author} {\bibfnamefont {H.~E.}\ \bibnamefont {Machlan}},\
	}\bibfield  {title} {\bibinfo {title} {The national bureau of standards
			atomic time scales: Generation, dissemination, stability, and accuracy},\
	}\href {https://doi.org/10.1109/TIM.1972.4314051} {\bibfield  {journal}
		{\bibinfo  {journal} {IEEE Transactions on Instrumentation and Measurement}\
		}\textbf {\bibinfo {volume} {21}},\ \bibinfo {pages} {388} (\bibinfo {year}
		{1972})}\BibitemShut {NoStop}%
	\bibitem [{\citenamefont {Walls}\ and\ \citenamefont
		{Allan}(1986)}]{Walls_1986}%
	\BibitemOpen
	\bibfield  {author} {\bibinfo {author} {\bibfnamefont {F.}~\bibnamefont
			{Walls}}\ and\ \bibinfo {author} {\bibfnamefont {D.}~\bibnamefont {Allan}},\
	}\bibfield  {title} {\bibinfo {title} {Measurements of frequency stability},\
	}\href {https://doi.org/10.1109/PROC.1986.13429} {\bibfield  {journal}
		{\bibinfo  {journal} {Proceedings of the IEEE}\ }\textbf {\bibinfo {volume}
			{74}},\ \bibinfo {pages} {162} (\bibinfo {year} {1986})}\BibitemShut
	{NoStop}%
	\bibitem [{\citenamefont {McCoy}\ and\ \citenamefont
		{Ernst}(1989)}]{McCoy_1989}%
	\BibitemOpen
	\bibfield  {author} {\bibinfo {author} {\bibfnamefont {M.}~\bibnamefont
			{McCoy}}\ and\ \bibinfo {author} {\bibfnamefont {R.}~\bibnamefont {Ernst}},\
	}\bibfield  {title} {\bibinfo {title} {Nuclear spin noise at room
			temperature},\ }\href {https://doi.org/10.1016/0009-2614(89)87537-2}
	{\bibfield  {journal} {\bibinfo  {journal} {Chemical Physics Letters}\
		}\textbf {\bibinfo {volume} {159}},\ \bibinfo {pages} {587} (\bibinfo {year}
		{1989})}\BibitemShut {NoStop}%
	\bibitem [{\citenamefont {Guéron}\ and\ \citenamefont
		{Leroy}(1989)}]{Gueron_1989}%
	\BibitemOpen
	\bibfield  {author} {\bibinfo {author} {\bibfnamefont {M.}~\bibnamefont
			{Guéron}}\ and\ \bibinfo {author} {\bibfnamefont {J.}~\bibnamefont
			{Leroy}},\ }\bibfield  {title} {\bibinfo {title} {Nmr of water protons. the
			detection of their nuclear-spin noise, and a simple determination of absolute
			probe sensitivity based on radiation damping},\ }\href
	{https://doi.org/10.1016/0022-2364(89)90338-7} {\bibfield  {journal}
		{\bibinfo  {journal} {Journal of Magnetic Resonance (1969)}\ }\textbf
		{\bibinfo {volume} {85}},\ \bibinfo {pages} {209} (\bibinfo {year}
		{1989})}\BibitemShut {NoStop}%
	\bibitem [{\citenamefont {Braginsky}\ and\ \citenamefont
		{Manukin}(1967)}]{Braginsky_1967}%
	\BibitemOpen
	\bibfield  {author} {\bibinfo {author} {\bibfnamefont {V.~B.}\ \bibnamefont
			{Braginsky}}\ and\ \bibinfo {author} {\bibfnamefont {A.}~\bibnamefont
			{Manukin}},\ }\bibfield  {title} {\bibinfo {title} {Ponderomotive effects of
			electromagnetic radiation},\ }\href
	{https://api.semanticscholar.org/CorpusID:118076715} {\bibfield  {journal}
		{\bibinfo  {journal} {Journal of Experimental and Theoretical Physics}\ }
		(\bibinfo {year} {1967})}\BibitemShut {NoStop}%
	\bibitem [{\citenamefont {Albrecht}\ \emph {et~al.}(1991)\citenamefont
		{Albrecht}, \citenamefont {Grütter}, \citenamefont {Horne},\ and\
		\citenamefont {Rugar}}]{Albrecht_1991}%
	\BibitemOpen
	\bibfield  {author} {\bibinfo {author} {\bibfnamefont {T.~R.}\ \bibnamefont
			{Albrecht}}, \bibinfo {author} {\bibfnamefont {P.}~\bibnamefont {Grütter}},
		\bibinfo {author} {\bibfnamefont {D.}~\bibnamefont {Horne}},\ and\ \bibinfo
		{author} {\bibfnamefont {D.}~\bibnamefont {Rugar}},\ }\bibfield  {title}
	{\bibinfo {title} {Frequency modulation detection using high-q cantilevers
			for enhanced force microscope sensitivity},\ }\href
	{https://doi.org/10.1063/1.347347} {\bibfield  {journal} {\bibinfo  {journal}
			{Journal of Applied Physics}\ }\textbf {\bibinfo {volume} {69}},\ \bibinfo
		{pages} {668} (\bibinfo {year} {1991})}\BibitemShut {NoStop}%
	\bibitem [{\citenamefont {Weber}\ \emph {et~al.}(2016)\citenamefont {Weber},
		\citenamefont {Güttinger}, \citenamefont {Noury}, \citenamefont
		{Vergara-Cruz},\ and\ \citenamefont {Bachtold}}]{Weber_2016}%
	\BibitemOpen
	\bibfield  {author} {\bibinfo {author} {\bibfnamefont {P.}~\bibnamefont
			{Weber}}, \bibinfo {author} {\bibfnamefont {J.}~\bibnamefont {Güttinger}},
		\bibinfo {author} {\bibfnamefont {A.}~\bibnamefont {Noury}}, \bibinfo
		{author} {\bibfnamefont {J.}~\bibnamefont {Vergara-Cruz}},\ and\ \bibinfo
		{author} {\bibfnamefont {A.}~\bibnamefont {Bachtold}},\ }\bibfield  {title}
	{\bibinfo {title} {Force sensitivity of multilayer graphene optomechanical
			devices},\ }\href {https://doi.org/10.1038/ncomms12496} {\bibfield  {journal}
		{\bibinfo  {journal} {Nature Communications}\ }\textbf {\bibinfo {volume}
			{7}},\ \bibinfo {pages} {12496} (\bibinfo {year} {2016})}\BibitemShut
	{NoStop}%
	\bibitem [{\citenamefont {Rugar}\ \emph {et~al.}(2004)\citenamefont {Rugar},
		\citenamefont {Budakian}, \citenamefont {Mamin},\ and\ \citenamefont
		{Chui}}]{Rugar_2004}%
	\BibitemOpen
	\bibfield  {author} {\bibinfo {author} {\bibfnamefont {D.}~\bibnamefont
			{Rugar}}, \bibinfo {author} {\bibfnamefont {R.}~\bibnamefont {Budakian}},
		\bibinfo {author} {\bibfnamefont {H.~J.}\ \bibnamefont {Mamin}},\ and\
		\bibinfo {author} {\bibfnamefont {B.~W.}\ \bibnamefont {Chui}},\ }\bibfield
	{title} {\bibinfo {title} {Single spin detection by magnetic resonance force
			microscopy},\ }\href {https://doi.org/10.1038/nature02658} {\bibfield
		{journal} {\bibinfo  {journal} {Nature}\ }\textbf {\bibinfo {volume} {430}},\
		\bibinfo {pages} {329} (\bibinfo {year} {2004})}\BibitemShut {NoStop}%
	\bibitem [{\citenamefont {Degen}\ \emph {et~al.}(2007)\citenamefont {Degen},
		\citenamefont {Poggio}, \citenamefont {Mamin},\ and\ \citenamefont
		{Rugar}}]{Degen_2007}%
	\BibitemOpen
	\bibfield  {author} {\bibinfo {author} {\bibfnamefont {C.~L.}\ \bibnamefont
			{Degen}}, \bibinfo {author} {\bibfnamefont {M.}~\bibnamefont {Poggio}},
		\bibinfo {author} {\bibfnamefont {H.~J.}\ \bibnamefont {Mamin}},\ and\
		\bibinfo {author} {\bibfnamefont {D.}~\bibnamefont {Rugar}},\ }\bibfield
	{title} {\bibinfo {title} {Role of spin noise in the detection of nanoscale
			ensembles of nuclear spins},\ }\href
	{https://doi.org/10.1103/PhysRevLett.99.250601} {\bibfield  {journal}
		{\bibinfo  {journal} {Physical Review Letters}\ }\textbf {\bibinfo {volume}
			{99}},\ \bibinfo {pages} {250601} (\bibinfo {year} {2007})}\BibitemShut
	{NoStop}%
	\bibitem [{\citenamefont {Herzog}\ \emph {et~al.}(2014)\citenamefont {Herzog},
		\citenamefont {Cadeddu}, \citenamefont {Xue}, \citenamefont {Peddibhotla},\
		and\ \citenamefont {Poggio}}]{Herzog_2014}%
	\BibitemOpen
	\bibfield  {author} {\bibinfo {author} {\bibfnamefont {B.~E.}\ \bibnamefont
			{Herzog}}, \bibinfo {author} {\bibfnamefont {D.}~\bibnamefont {Cadeddu}},
		\bibinfo {author} {\bibfnamefont {F.}~\bibnamefont {Xue}}, \bibinfo {author}
		{\bibfnamefont {P.}~\bibnamefont {Peddibhotla}},\ and\ \bibinfo {author}
		{\bibfnamefont {M.}~\bibnamefont {Poggio}},\ }\bibfield  {title} {\bibinfo
		{title} {{Boundary between the thermal and statistical polarization regimes
				in a nuclear spin ensemble}},\ }\href {https://doi.org/10.1063/1.4892361}
	{\bibfield  {journal} {\bibinfo  {journal} {Applied Physics Letters}\
		}\textbf {\bibinfo {volume} {105}},\ \bibinfo {pages} {043112} (\bibinfo
		{year} {2014})}\BibitemShut {NoStop}%
	\bibitem [{\citenamefont {Kubo}(1962)}]{Kubo_1962}%
	\BibitemOpen
	\bibfield  {author} {\bibinfo {author} {\bibfnamefont {R.}~\bibnamefont
			{Kubo}},\ }\bibfield  {title} {\bibinfo {title} {Generalized cumulant
			expansion method},\ }\href {https://doi.org/10.1143/JPSJ.17.1100} {\bibfield
		{journal} {\bibinfo  {journal} {Journal of the Physical Society of Japan}\
		}\textbf {\bibinfo {volume} {17}},\ \bibinfo {pages} {1100} (\bibinfo {year}
		{1962})}\BibitemShut {NoStop}%
	\bibitem [{\citenamefont {Poggio}\ \emph {et~al.}(2007)\citenamefont {Poggio},
		\citenamefont {Degen}, \citenamefont {Rettner}, \citenamefont {Mamin},\ and\
		\citenamefont {Rugar}}]{Poggio_2007}%
	\BibitemOpen
	\bibfield  {author} {\bibinfo {author} {\bibfnamefont {M.}~\bibnamefont
			{Poggio}}, \bibinfo {author} {\bibfnamefont {C.~L.}\ \bibnamefont {Degen}},
		\bibinfo {author} {\bibfnamefont {C.~T.}\ \bibnamefont {Rettner}}, \bibinfo
		{author} {\bibfnamefont {H.~J.}\ \bibnamefont {Mamin}},\ and\ \bibinfo
		{author} {\bibfnamefont {D.}~\bibnamefont {Rugar}},\ }\bibfield  {title}
	{\bibinfo {title} {Nuclear magnetic resonance force microscopy with a
			microwire rf source},\ }\href {https://doi.org/10.1063/1.2752536} {\bibfield
		{journal} {\bibinfo  {journal} {Applied Physics Letters}\ }\textbf {\bibinfo
			{volume} {90}},\ \bibinfo {pages} {263111} (\bibinfo {year}
		{2007})}\BibitemShut {NoStop}%
	\bibitem [{\citenamefont {Carver}\ and\ \citenamefont
		{Slichter}(1953)}]{Carver_1953}%
	\BibitemOpen
	\bibfield  {author} {\bibinfo {author} {\bibfnamefont {T.~R.}\ \bibnamefont
			{Carver}}\ and\ \bibinfo {author} {\bibfnamefont {C.~P.}\ \bibnamefont
			{Slichter}},\ }\bibfield  {title} {\bibinfo {title} {Polarization of nuclear
			spins in metals},\ }\href {https://doi.org/10.1103/PhysRev.92.212.2}
	{\bibfield  {journal} {\bibinfo  {journal} {Physical Review}\ }\textbf
		{\bibinfo {volume} {92}},\ \bibinfo {pages} {212} (\bibinfo {year}
		{1953})}\BibitemShut {NoStop}%
	\bibitem [{\citenamefont {Carver}\ and\ \citenamefont
		{Slichter}(1956)}]{Carver_1956}%
	\BibitemOpen
	\bibfield  {author} {\bibinfo {author} {\bibfnamefont {T.~R.}\ \bibnamefont
			{Carver}}\ and\ \bibinfo {author} {\bibfnamefont {C.~P.}\ \bibnamefont
			{Slichter}},\ }\bibfield  {title} {\bibinfo {title} {Experimental
			verification of the overhauser nuclear polarization effect},\ }\href
	{https://doi.org/10.1103/PhysRev.102.975} {\bibfield  {journal} {\bibinfo
			{journal} {Physical Review}\ }\textbf {\bibinfo {volume} {102}},\ \bibinfo
		{pages} {975} (\bibinfo {year} {1956})}\BibitemShut {NoStop}%
	\bibitem [{\citenamefont {Hunger}\ \emph {et~al.}(2011)\citenamefont {Hunger},
		\citenamefont {Camerer}, \citenamefont {Korppi}, \citenamefont {Jöckel},
		\citenamefont {Hänsch},\ and\ \citenamefont {Treutlein}}]{Hunger_2011}%
	\BibitemOpen
	\bibfield  {author} {\bibinfo {author} {\bibfnamefont {D.}~\bibnamefont
			{Hunger}}, \bibinfo {author} {\bibfnamefont {S.}~\bibnamefont {Camerer}},
		\bibinfo {author} {\bibfnamefont {M.}~\bibnamefont {Korppi}}, \bibinfo
		{author} {\bibfnamefont {A.}~\bibnamefont {Jöckel}}, \bibinfo {author}
		{\bibfnamefont {T.}~\bibnamefont {Hänsch}},\ and\ \bibinfo {author}
		{\bibfnamefont {P.}~\bibnamefont {Treutlein}},\ }\bibfield  {title} {\bibinfo
		{title} {Coupling ultracold atoms to mechanical oscillators},\ }\href
	{https://doi.org/10.1016/j.crhy.2011.04.015} {\bibfield  {journal} {\bibinfo
			{journal} {Comptes Rendus Physique}\ }\textbf {\bibinfo {volume} {12}},\
		\bibinfo {pages} {871} (\bibinfo {year} {2011})}\BibitemShut {NoStop}%
	\bibitem [{\citenamefont {Butler}\ and\ \citenamefont
		{Weitekamp}(2011)}]{Butler_2011}%
	\BibitemOpen
	\bibfield  {author} {\bibinfo {author} {\bibfnamefont {M.~C.}\ \bibnamefont
			{Butler}}\ and\ \bibinfo {author} {\bibfnamefont {D.~P.}\ \bibnamefont
			{Weitekamp}},\ }\bibfield  {title} {\bibinfo {title} {Polarization of nuclear
			spins by a cold nanoscale resonator},\ }\href
	{https://doi.org/10.1103/PhysRevA.84.063407} {\bibfield  {journal} {\bibinfo
			{journal} {Physical Review A}\ }\textbf {\bibinfo {volume} {84}},\ \bibinfo
		{pages} {063407} (\bibinfo {year} {2011})}\BibitemShut {NoStop}%
	\bibitem [{\citenamefont {Nunnenkamp}\ \emph {et~al.}(2014)\citenamefont
		{Nunnenkamp}, \citenamefont {Sudhir}, \citenamefont {Feofanov}, \citenamefont
		{Roulet},\ and\ \citenamefont {Kippenberg}}]{Nunnenkamp_2014}%
	\BibitemOpen
	\bibfield  {author} {\bibinfo {author} {\bibfnamefont {A.}~\bibnamefont
			{Nunnenkamp}}, \bibinfo {author} {\bibfnamefont {V.}~\bibnamefont {Sudhir}},
		\bibinfo {author} {\bibfnamefont {A.}~\bibnamefont {Feofanov}}, \bibinfo
		{author} {\bibfnamefont {A.}~\bibnamefont {Roulet}},\ and\ \bibinfo {author}
		{\bibfnamefont {T.}~\bibnamefont {Kippenberg}},\ }\bibfield  {title}
	{\bibinfo {title} {Quantum-limited amplification and parametric instability
			in the reversed dissipation regime of cavity optomechanics},\ }\href
	{https://doi.org/10.1103/PhysRevLett.113.023604} {\bibfield  {journal}
		{\bibinfo  {journal} {Physical Review Letters}\ }\textbf {\bibinfo {volume}
			{113}},\ \bibinfo {pages} {023604} (\bibinfo {year} {2014})}\BibitemShut
	{NoStop}%
	\bibitem [{\citenamefont {Ohta}\ \emph {et~al.}(2021)\citenamefont {Ohta},
		\citenamefont {Herpin}, \citenamefont {Bastidas}, \citenamefont {Tawara},
		\citenamefont {Yamaguchi},\ and\ \citenamefont {Okamoto}}]{Ohta_2021}%
	\BibitemOpen
	\bibfield  {author} {\bibinfo {author} {\bibfnamefont {R.}~\bibnamefont
			{Ohta}}, \bibinfo {author} {\bibfnamefont {L.}~\bibnamefont {Herpin}},
		\bibinfo {author} {\bibfnamefont {V.}~\bibnamefont {Bastidas}}, \bibinfo
		{author} {\bibfnamefont {T.}~\bibnamefont {Tawara}}, \bibinfo {author}
		{\bibfnamefont {H.}~\bibnamefont {Yamaguchi}},\ and\ \bibinfo {author}
		{\bibfnamefont {H.}~\bibnamefont {Okamoto}},\ }\bibfield  {title} {\bibinfo
		{title} {Rare-earth-mediated optomechanical system in the reversed
			dissipation regime},\ }\href {https://doi.org/10.1103/PhysRevLett.126.047404}
	{\bibfield  {journal} {\bibinfo  {journal} {Physical Review Letters}\
		}\textbf {\bibinfo {volume} {126}},\ \bibinfo {pages} {047404} (\bibinfo
		{year} {2021})}\BibitemShut {NoStop}%
	\bibitem [{\citenamefont {Kirton}\ and\ \citenamefont
		{Keeling}(2017)}]{Kirton_2017}%
	\BibitemOpen
	\bibfield  {author} {\bibinfo {author} {\bibfnamefont {P.}~\bibnamefont
			{Kirton}}\ and\ \bibinfo {author} {\bibfnamefont {J.}~\bibnamefont
			{Keeling}},\ }\bibfield  {title} {\bibinfo {title} {Suppressing and restoring
			the dicke superradiance transition by dephasing and decay},\ }\href
	{https://doi.org/10.1103/PhysRevLett.118.123602} {\bibfield  {journal}
		{\bibinfo  {journal} {Physical Review Letters}\ }\textbf {\bibinfo {volume}
			{118}},\ \bibinfo {pages} {123602} (\bibinfo {year} {2017})}\BibitemShut
	{NoStop}%
	\bibitem [{\citenamefont {Dalla~Torre}\ \emph {et~al.}(2016)\citenamefont
		{Dalla~Torre}, \citenamefont {Shchadilova}, \citenamefont {Wilner},
		\citenamefont {Lukin},\ and\ \citenamefont {Demler}}]{Dalla_Torre_2016}%
	\BibitemOpen
	\bibfield  {author} {\bibinfo {author} {\bibfnamefont {E.~G.}\ \bibnamefont
			{Dalla~Torre}}, \bibinfo {author} {\bibfnamefont {Y.}~\bibnamefont
			{Shchadilova}}, \bibinfo {author} {\bibfnamefont {E.~Y.}\ \bibnamefont
			{Wilner}}, \bibinfo {author} {\bibfnamefont {M.~D.}\ \bibnamefont {Lukin}},\
		and\ \bibinfo {author} {\bibfnamefont {E.}~\bibnamefont {Demler}},\
	}\bibfield  {title} {\bibinfo {title} {Dicke phase transition without total
			spin conservation},\ }\href {https://doi.org/10.1103/PhysRevA.94.061802}
	{\bibfield  {journal} {\bibinfo  {journal} {Physical Review A}\ }\textbf
		{\bibinfo {volume} {94}},\ \bibinfo {pages} {061802} (\bibinfo {year}
		{2016})}\BibitemShut {NoStop}%
	\bibitem [{\citenamefont {Dobrindt}\ and\ \citenamefont
		{Kippenberg}(2010)}]{Dobrindt_2010}%
	\BibitemOpen
	\bibfield  {author} {\bibinfo {author} {\bibfnamefont {J.~M.}\ \bibnamefont
			{Dobrindt}}\ and\ \bibinfo {author} {\bibfnamefont {T.~J.}\ \bibnamefont
			{Kippenberg}},\ }\bibfield  {title} {\bibinfo {title} {Theoretical analysis
			of mechanical displacement measurement using a multiple cavity mode
			transducer},\ }\href {https://doi.org/10.1103/PhysRevLett.104.033901}
	{\bibfield  {journal} {\bibinfo  {journal} {Physical Review Letters}\
		}\textbf {\bibinfo {volume} {104}},\ \bibinfo {pages} {033901} (\bibinfo
		{year} {2010})}\BibitemShut {NoStop}%
	\bibitem [{\citenamefont {Burgwal}\ \emph {et~al.}(2020)\citenamefont
		{Burgwal}, \citenamefont {Pino},\ and\ \citenamefont
		{Verhagen}}]{Burgwal_2020}%
	\BibitemOpen
	\bibfield  {author} {\bibinfo {author} {\bibfnamefont {R.}~\bibnamefont
			{Burgwal}}, \bibinfo {author} {\bibfnamefont {J.~d.}\ \bibnamefont {Pino}},\
		and\ \bibinfo {author} {\bibfnamefont {E.}~\bibnamefont {Verhagen}},\
	}\bibfield  {title} {\bibinfo {title} {Comparing nonlinear optomechanical
			coupling in membrane-in-the-middle and single-cavity systems},\ }\href
	{https://doi.org/10.1088/1367-2630/abc1c8} {\bibfield  {journal} {\bibinfo
			{journal} {New Journal of Physics}\ }\textbf {\bibinfo {volume} {22}},\
		\bibinfo {pages} {113006} (\bibinfo {year} {2020})}\BibitemShut {NoStop}%
	\bibitem [{\citenamefont {Burgwal}\ and\ \citenamefont
		{Verhagen}(2023)}]{Burgwal_2023}%
	\BibitemOpen
	\bibfield  {author} {\bibinfo {author} {\bibfnamefont {R.}~\bibnamefont
			{Burgwal}}\ and\ \bibinfo {author} {\bibfnamefont {E.}~\bibnamefont
			{Verhagen}},\ }\bibfield  {title} {\bibinfo {title} {Enhanced nonlinear
			optomechanics in a coupled-mode photonic crystal device},\ }\href
	{https://doi.org/10.1038/s41467-023-37138-z} {\bibfield  {journal} {\bibinfo
			{journal} {Nature Communications}\ }\textbf {\bibinfo {volume} {14}},\
		\bibinfo {pages} {1526} (\bibinfo {year} {2023})}\BibitemShut {NoStop}%
\end{thebibliography}

\begin{thebibliography}{19}%
	\makeatletter
	\providecommand \@ifxundefined [1]{%
		\@ifx{#1\undefined}
	}%
	\providecommand \@ifnum [1]{%
		\ifnum #1\expandafter \@firstoftwo
		\else \expandafter \@secondoftwo
		\fi
	}%
	\providecommand \@ifx [1]{%
		\ifx #1\expandafter \@firstoftwo
		\else \expandafter \@secondoftwo
		\fi
	}%
	\providecommand \natexlab [1]{#1}%
	\providecommand \enquote  [1]{``#1''}%
	\providecommand \bibnamefont  [1]{#1}%
	\providecommand \bibfnamefont [1]{#1}%
	\providecommand \citenamefont [1]{#1}%
	\providecommand \href@noop [0]{\@secondoftwo}%
	\providecommand \href [0]{\begingroup \@sanitize@url \@href}%
	\providecommand \@href[1]{\@@startlink{#1}\@@href}%
	\providecommand \@@href[1]{\endgroup#1\@@endlink}%
	\providecommand \@sanitize@url [0]{\catcode `\\12\catcode `\$12\catcode
		`\&12\catcode `\#12\catcode `\^12\catcode `\_12\catcode `\%12\relax}%
	\providecommand \@@startlink[1]{}%
	\providecommand \@@endlink[0]{}%
	\providecommand \url  [0]{\begingroup\@sanitize@url \@url }%
	\providecommand \@url [1]{\endgroup\@href {#1}{\urlprefix }}%
	\providecommand \urlprefix  [0]{URL }%
	\providecommand \Eprint [0]{\href }%
	\providecommand \doibase [0]{https://doi.org/}%
	\providecommand \selectlanguage [0]{\@gobble}%
	\providecommand \bibinfo  [0]{\@secondoftwo}%
	\providecommand \bibfield  [0]{\@secondoftwo}%
	\providecommand \translation [1]{[#1]}%
	\providecommand \BibitemOpen [0]{}%
	\providecommand \bibitemStop [0]{}%
	\providecommand \bibitemNoStop [0]{.\EOS\space}%
	\providecommand \EOS [0]{\spacefactor3000\relax}%
	\providecommand \BibitemShut  [1]{\csname bibitem#1\endcsname}%
	\let\auto@bib@innerbib\@empty
	\bibitem [{\citenamefont {Niinikoski}(2020)}]{Niinikoski_2020}%
	\BibitemOpen
	\bibfield  {author} {\bibinfo {author} {\bibfnamefont {T.~O.}\ \bibnamefont
			{Niinikoski}},\ }\href {https://doi.org/10.1017/9781108567435} {\emph
		{\bibinfo {title} {The Physics of Polarized Targets}}}\ (\bibinfo
	{publisher} {Cambridge University Press},\ \bibinfo {year}
	{2020})\BibitemShut {NoStop}%
	\bibitem [{\citenamefont {Rand}(2005)}]{Rand_2005}%
	\BibitemOpen
	\bibfield  {author} {\bibinfo {author} {\bibfnamefont {R.~H.}\ \bibnamefont
			{Rand}},\ }\href@noop {} {\bibinfo {title} {{Lecture Notes on Nonlinear
				Vibrations}}},\ \bibinfo {howpublished}
	{\url{http://audiophile.tam.cornell.edu/randdocs/nlvibe52.pdf}} (\bibinfo
	{year} {2005})\BibitemShut {NoStop}%
	\bibitem [{\citenamefont {Krack}\ and\ \citenamefont
		{Gross}(2019)}]{Krack_2019}%
	\BibitemOpen
	\bibfield  {author} {\bibinfo {author} {\bibfnamefont {M.}~\bibnamefont
			{Krack}}\ and\ \bibinfo {author} {\bibfnamefont {J.}~\bibnamefont {Gross}},\
	}\href {https://doi.org/10.1007/978-3-030-14023-6} {\emph {\bibinfo {title}
			{Harmonic Balance for Nonlinear Vibration Problems}}}\ (\bibinfo  {publisher}
	{Springer International Publishing},\ \bibinfo {year} {2019})\BibitemShut
	{NoStop}%
	\bibitem [{\citenamefont {Breiding}\ and\ \citenamefont
		{Timme}(2018)}]{Breiding_2018}%
	\BibitemOpen
	\bibfield  {author} {\bibinfo {author} {\bibfnamefont {P.}~\bibnamefont
			{Breiding}}\ and\ \bibinfo {author} {\bibfnamefont {S.}~\bibnamefont
			{Timme}},\ }\bibfield  {title} {\bibinfo {title} {Homotopycontinuation.jl: A
			package for homotopy continuation in julia},\ }in\ \href
	{https://doi.org/10.1007/978-3-319-96418-8_54} {\emph {\bibinfo {booktitle}
			{Lecture Notes in Computer Science}}},\ Vol.\ \bibinfo {volume} {10931 LNCS}\
	(\bibinfo  {publisher} {Springer International Publishing},\ \bibinfo {year}
	{2018})\ pp.\ \bibinfo {pages} {458--465}\BibitemShut {NoStop}%
	\bibitem [{\citenamefont {Košata}\ \emph {et~al.}(2022)\citenamefont
		{Košata}, \citenamefont {del Pino}, \citenamefont {Heugel},\ and\
		\citenamefont {Zilberberg}}]{Kosata_2022_SP}%
	\BibitemOpen
	\bibfield  {author} {\bibinfo {author} {\bibfnamefont {J.}~\bibnamefont
			{Košata}}, \bibinfo {author} {\bibfnamefont {J.}~\bibnamefont {del Pino}},
		\bibinfo {author} {\bibfnamefont {T.~L.}\ \bibnamefont {Heugel}},\ and\
		\bibinfo {author} {\bibfnamefont {O.}~\bibnamefont {Zilberberg}},\ }\bibfield
	{title} {\bibinfo {title} {Harmonicbalance.jl: A julia suite for nonlinear
			dynamics using harmonic balance},\ }\href
	{https://doi.org/10.21468/SciPostPhysCodeb.6} {\bibfield  {journal} {\bibinfo
			{journal} {SciPost Physics Codebases}\ ,\ \bibinfo {pages} {6}} (\bibinfo
		{year} {2022})}\BibitemShut {NoStop}%
	\bibitem [{\citenamefont {Lugiato}\ \emph {et~al.}(1984)\citenamefont
		{Lugiato}, \citenamefont {Mandel},\ and\ \citenamefont
		{Narducci}}]{Lugiato_1984}%
	\BibitemOpen
	\bibfield  {author} {\bibinfo {author} {\bibfnamefont {L.~A.}\ \bibnamefont
			{Lugiato}}, \bibinfo {author} {\bibfnamefont {P.}~\bibnamefont {Mandel}},\
		and\ \bibinfo {author} {\bibfnamefont {L.~M.}\ \bibnamefont {Narducci}},\
	}\bibfield  {title} {\bibinfo {title} {Adiabatic elimination in nonlinear
			dynamical systems},\ }\href {https://doi.org/10.1103/PhysRevA.29.1438}
	{\bibfield  {journal} {\bibinfo  {journal} {Physical Review A}\ }\textbf
		{\bibinfo {volume} {29}},\ \bibinfo {pages} {1438} (\bibinfo {year}
		{1984})}\BibitemShut {NoStop}%
	\bibitem [{\citenamefont {Bhaseen}\ \emph {et~al.}(2012)\citenamefont
		{Bhaseen}, \citenamefont {Mayoh}, \citenamefont {Simons},\ and\ \citenamefont
		{Keeling}}]{Bhaseen_2012}%
	\BibitemOpen
	\bibfield  {author} {\bibinfo {author} {\bibfnamefont {M.~J.}\ \bibnamefont
			{Bhaseen}}, \bibinfo {author} {\bibfnamefont {J.}~\bibnamefont {Mayoh}},
		\bibinfo {author} {\bibfnamefont {B.~D.}\ \bibnamefont {Simons}},\ and\
		\bibinfo {author} {\bibfnamefont {J.}~\bibnamefont {Keeling}},\ }\bibfield
	{title} {\bibinfo {title} {Dynamics of nonequilibrium dicke models},\ }\href
	{https://doi.org/10.1103/PhysRevA.85.013817} {\bibfield  {journal} {\bibinfo
			{journal} {Physical Review A}\ }\textbf {\bibinfo {volume} {85}},\ \bibinfo
		{pages} {013817} (\bibinfo {year} {2012})}\BibitemShut {NoStop}%
	\bibitem [{\citenamefont {Chitra}\ and\ \citenamefont
		{Zilberberg}(2015)}]{Chitra_2015}%
	\BibitemOpen
	\bibfield  {author} {\bibinfo {author} {\bibfnamefont {R.}~\bibnamefont
			{Chitra}}\ and\ \bibinfo {author} {\bibfnamefont {O.}~\bibnamefont
			{Zilberberg}},\ }\bibfield  {title} {\bibinfo {title} {Dynamical many-body
			phases of the parametrically driven, dissipative dicke model},\ }\href
	{https://doi.org/10.1103/PhysRevA.92.023815} {\bibfield  {journal} {\bibinfo
			{journal} {Physical Review A}\ }\textbf {\bibinfo {volume} {92}},\ \bibinfo
		{pages} {023815} (\bibinfo {year} {2015})}\BibitemShut {NoStop}%
	\bibitem [{\citenamefont {Tsaturyan}\ \emph {et~al.}(2017)\citenamefont
		{Tsaturyan}, \citenamefont {Barg}, \citenamefont {Polzik},\ and\
		\citenamefont {Schliesser}}]{Tsaturyan_2017}%
	\BibitemOpen
	\bibfield  {author} {\bibinfo {author} {\bibfnamefont {Y.}~\bibnamefont
			{Tsaturyan}}, \bibinfo {author} {\bibfnamefont {A.}~\bibnamefont {Barg}},
		\bibinfo {author} {\bibfnamefont {E.~S.}\ \bibnamefont {Polzik}},\ and\
		\bibinfo {author} {\bibfnamefont {A.}~\bibnamefont {Schliesser}},\ }\bibfield
	{title} {\bibinfo {title} {Ultracoherent nanomechanical resonators via soft
			clamping and dissipation dilution},\ }\href
	{https://doi.org/10.1038/nnano.2017.101} {\bibfield  {journal} {\bibinfo
			{journal} {Nature Nanotechnology}\ }\textbf {\bibinfo {volume} {12}},\
		\bibinfo {pages} {776} (\bibinfo {year} {2017})}\BibitemShut {NoStop}%
	\bibitem [{\citenamefont {Aspelmeyer}\ \emph {et~al.}(2014)\citenamefont
		{Aspelmeyer}, \citenamefont {Kippenberg},\ and\ \citenamefont
		{Marquardt}}]{Aspelmeyer_2014}%
	\BibitemOpen
	\bibfield  {author} {\bibinfo {author} {\bibfnamefont {M.}~\bibnamefont
			{Aspelmeyer}}, \bibinfo {author} {\bibfnamefont {T.~J.}\ \bibnamefont
			{Kippenberg}},\ and\ \bibinfo {author} {\bibfnamefont {F.}~\bibnamefont
			{Marquardt}},\ }\bibfield  {title} {\bibinfo {title} {Cavity optomechanics},\
	}\href {https://doi.org/10.1103/RevModPhys.86.1391} {\bibfield  {journal}
		{\bibinfo  {journal} {Reviews of Modern Physics}\ }\textbf {\bibinfo {volume}
			{86}},\ \bibinfo {pages} {1391} (\bibinfo {year} {2014})}\BibitemShut
	{NoStop}%
	\bibitem [{\citenamefont {Greenberg}\ \emph {et~al.}(2009)\citenamefont
		{Greenberg}, \citenamefont {Il’ichev},\ and\ \citenamefont
		{Nori}}]{Greenberg_2009}%
	\BibitemOpen
	\bibfield  {author} {\bibinfo {author} {\bibfnamefont {Y.~S.}\ \bibnamefont
			{Greenberg}}, \bibinfo {author} {\bibfnamefont {E.}~\bibnamefont
			{Il’ichev}},\ and\ \bibinfo {author} {\bibfnamefont {F.}~\bibnamefont
			{Nori}},\ }\bibfield  {title} {\bibinfo {title} {Cooling a magnetic resonance
			force microscope via the dynamical back action of nuclear spins},\ }\href
	{https://doi.org/10.1103/PhysRevB.80.214423} {\bibfield  {journal} {\bibinfo
			{journal} {Physical Review B}\ }\textbf {\bibinfo {volume} {80}},\ \bibinfo
		{pages} {214423} (\bibinfo {year} {2009})}\BibitemShut {NoStop}%
	\bibitem [{\citenamefont {Ghadimi}\ \emph {et~al.}(2018)\citenamefont
		{Ghadimi}, \citenamefont {Fedorov}, \citenamefont {Engelsen}, \citenamefont
		{Bereyhi}, \citenamefont {Schilling}, \citenamefont {Wilson},\ and\
		\citenamefont {Kippenberg}}]{Ghadimi_2018}%
	\BibitemOpen
	\bibfield  {author} {\bibinfo {author} {\bibfnamefont {A.~H.}\ \bibnamefont
			{Ghadimi}}, \bibinfo {author} {\bibfnamefont {S.~A.}\ \bibnamefont
			{Fedorov}}, \bibinfo {author} {\bibfnamefont {N.~J.}\ \bibnamefont
			{Engelsen}}, \bibinfo {author} {\bibfnamefont {M.~J.}\ \bibnamefont
			{Bereyhi}}, \bibinfo {author} {\bibfnamefont {R.}~\bibnamefont {Schilling}},
		\bibinfo {author} {\bibfnamefont {D.~J.}\ \bibnamefont {Wilson}},\ and\
		\bibinfo {author} {\bibfnamefont {T.~J.}\ \bibnamefont {Kippenberg}},\
	}\bibfield  {title} {\bibinfo {title} {Elastic strain engineering for
			ultralow mechanical dissipation},\ }\href
	{https://doi.org/10.1126/science.aar6939} {\bibfield  {journal} {\bibinfo
			{journal} {Science}\ }\textbf {\bibinfo {volume} {360}},\ \bibinfo {pages}
		{764} (\bibinfo {year} {2018})}\BibitemShut {NoStop}%
	\bibitem [{\citenamefont {Weber}\ \emph {et~al.}(2016)\citenamefont {Weber},
		\citenamefont {Güttinger}, \citenamefont {Noury}, \citenamefont
		{Vergara-Cruz},\ and\ \citenamefont {Bachtold}}]{Weber_2016}%
	\BibitemOpen
	\bibfield  {author} {\bibinfo {author} {\bibfnamefont {P.}~\bibnamefont
			{Weber}}, \bibinfo {author} {\bibfnamefont {J.}~\bibnamefont {Güttinger}},
		\bibinfo {author} {\bibfnamefont {A.}~\bibnamefont {Noury}}, \bibinfo
		{author} {\bibfnamefont {J.}~\bibnamefont {Vergara-Cruz}},\ and\ \bibinfo
		{author} {\bibfnamefont {A.}~\bibnamefont {Bachtold}},\ }\bibfield  {title}
	{\bibinfo {title} {Force sensitivity of multilayer graphene optomechanical
			devices},\ }\href {https://doi.org/10.1038/ncomms12496} {\bibfield  {journal}
		{\bibinfo  {journal} {Nature Communications}\ }\textbf {\bibinfo {volume}
			{7}},\ \bibinfo {pages} {12496} (\bibinfo {year} {2016})}\BibitemShut
	{NoStop}%
	\bibitem [{\citenamefont {Kocman}\ \emph {et~al.}(2019)\citenamefont {Kocman},
		\citenamefont {Di~Mauro}, \citenamefont {Veglia},\ and\ \citenamefont
		{Ramamoorthy}}]{Kocman_2019}%
	\BibitemOpen
	\bibfield  {author} {\bibinfo {author} {\bibfnamefont {V.}~\bibnamefont
			{Kocman}}, \bibinfo {author} {\bibfnamefont {G.~M.}\ \bibnamefont
			{Di~Mauro}}, \bibinfo {author} {\bibfnamefont {G.}~\bibnamefont {Veglia}},\
		and\ \bibinfo {author} {\bibfnamefont {A.}~\bibnamefont {Ramamoorthy}},\
	}\bibfield  {title} {\bibinfo {title} {Use of paramagnetic systems to
			speed-up nmr data acquisition and for structural and dynamic studies},\
	}\href {https://doi.org/10.1016/j.ssnmr.2019.07.002} {\bibfield  {journal}
		{\bibinfo  {journal} {Solid State Nuclear Magnetic Resonance}\ }\textbf
		{\bibinfo {volume} {102}},\ \bibinfo {pages} {36} (\bibinfo {year}
		{2019})}\BibitemShut {NoStop}%
	\bibitem [{\citenamefont {Bloch}(1946)}]{Bloch_1946}%
	\BibitemOpen
	\bibfield  {author} {\bibinfo {author} {\bibfnamefont {F.}~\bibnamefont
			{Bloch}},\ }\bibfield  {title} {\bibinfo {title} {Nuclear induction},\ }\href
	{https://doi.org/10.1103/PhysRev.70.460} {\bibfield  {journal} {\bibinfo
			{journal} {Physical Review}\ }\textbf {\bibinfo {volume} {70}},\ \bibinfo
		{pages} {460} (\bibinfo {year} {1946})}\BibitemShut {NoStop}%
	\bibitem [{\citenamefont {Degen}\ \emph {et~al.}(2007)\citenamefont {Degen},
		\citenamefont {Poggio}, \citenamefont {Mamin},\ and\ \citenamefont
		{Rugar}}]{Degen_2007}%
	\BibitemOpen
	\bibfield  {author} {\bibinfo {author} {\bibfnamefont {C.~L.}\ \bibnamefont
			{Degen}}, \bibinfo {author} {\bibfnamefont {M.}~\bibnamefont {Poggio}},
		\bibinfo {author} {\bibfnamefont {H.~J.}\ \bibnamefont {Mamin}},\ and\
		\bibinfo {author} {\bibfnamefont {D.}~\bibnamefont {Rugar}},\ }\bibfield
	{title} {\bibinfo {title} {Role of spin noise in the detection of nanoscale
			ensembles of nuclear spins},\ }\href
	{https://doi.org/10.1103/PhysRevLett.99.250601} {\bibfield  {journal}
		{\bibinfo  {journal} {Physical Review Letters}\ }\textbf {\bibinfo {volume}
			{99}},\ \bibinfo {pages} {250601} (\bibinfo {year} {2007})}\BibitemShut
	{NoStop}%
	\bibitem [{\citenamefont {Hairer}\ \emph {et~al.}(1993)\citenamefont {Hairer},
		\citenamefont {Wanner},\ and\ \citenamefont {Nørsett}}]{Hairer_1993}%
	\BibitemOpen
	\bibfield  {author} {\bibinfo {author} {\bibfnamefont {E.}~\bibnamefont
			{Hairer}}, \bibinfo {author} {\bibfnamefont {G.}~\bibnamefont {Wanner}},\
		and\ \bibinfo {author} {\bibfnamefont {S.~P.}\ \bibnamefont {Nørsett}},\
	}\href {https://doi.org/10.1007/978-3-540-78862-1} {\emph {\bibinfo {title}
			{Solving Ordinary Differential Equations I}}}\ (\bibinfo  {publisher}
	{Springer Berlin Heidelberg},\ \bibinfo {year} {1993})\BibitemShut {NoStop}%
	\bibitem [{\citenamefont {{Krass, Marc-Dominik}}(2022)}]{Krass_2022}%
	\BibitemOpen
	\bibfield  {author} {\bibinfo {author} {\bibnamefont {{Krass,
					Marc-Dominik}}},\ }\emph {\bibinfo {title} {3D Magnetic Resonance Force
			Microscopy}},\ \href {https://doi.org/10.3929/ethz-b-000536378} {Ph.D.
		thesis},\ \bibinfo  {school} {ETH Zurich} (\bibinfo {year}
	{2022})\BibitemShut {NoStop}%
	\bibitem [{\citenamefont {Longenecker}\ \emph {et~al.}(2012)\citenamefont
		{Longenecker}, \citenamefont {Mamin}, \citenamefont {Senko}, \citenamefont
		{Chen}, \citenamefont {Rettner}, \citenamefont {Rugar},\ and\ \citenamefont
		{Marohn}}]{Longenecker_2012}%
	\BibitemOpen
	\bibfield  {author} {\bibinfo {author} {\bibfnamefont {J.~G.}\ \bibnamefont
			{Longenecker}}, \bibinfo {author} {\bibfnamefont {H.~J.}\ \bibnamefont
			{Mamin}}, \bibinfo {author} {\bibfnamefont {A.~W.}\ \bibnamefont {Senko}},
		\bibinfo {author} {\bibfnamefont {L.}~\bibnamefont {Chen}}, \bibinfo {author}
		{\bibfnamefont {C.~T.}\ \bibnamefont {Rettner}}, \bibinfo {author}
		{\bibfnamefont {D.}~\bibnamefont {Rugar}},\ and\ \bibinfo {author}
		{\bibfnamefont {J.~A.}\ \bibnamefont {Marohn}},\ }\bibfield  {title}
	{\bibinfo {title} {High-gradient nanomagnets on cantilevers for sensitive
			detection of nuclear magnetic resonance},\ }\href
	{https://doi.org/10.1021/nn3030628} {\bibfield  {journal} {\bibinfo
			{journal} {ACS Nano}\ }\textbf {\bibinfo {volume} {6}},\ \bibinfo {pages}
		{9637} (\bibinfo {year} {2012})}\BibitemShut {NoStop}%
\end{thebibliography}
\end{document}